\documentclass{article}

\usepackage{amsmath,amsfonts,stmaryrd,amssymb} % Math packages
\usepackage[skip=1em,font=footnotesize]{caption}
\usepackage{newpxtext}
\usepackage{fourier}

\usepackage[utf8]{inputenc}
\usepackage{float}
\usepackage{graphicx}
\usepackage{xcolor}
\usepackage{soul}
\usepackage{booktabs}

\usepackage{authblk}
\usepackage[backend=biber,style=authoryear, maxcitenames=2]{biblatex}
\usepackage{xurl}
\usepackage{doi}
\usepackage{physics}
\usepackage{xfrac}

\usepackage{titlesec}
\usepackage{mathtools}

\makeatletter
\renewcommand{\maketitle}{\bgroup\setlength{\parindent}{-2pt}
\begin{flushleft}
  \textbf{\Large\@title}

  \@author
\end{flushleft}\egroup
}
\makeatother

\titleformat*{\section}{\large\bfseries}
\titleformat*{\subsection}{\normalsize\bfseries}
\titleformat*{\subsubsection}{\large\bfseries}
\titleformat*{\paragraph}{\large\bfseries}
\titleformat*{\subparagraph}{\large\bfseries}

\usepackage{enumerate} % Custom item numbers for enumerations

\usepackage{geometry} % Required for adjusting page dimensions and margins

\usepackage[utf8]{inputenc} % Required for inputting international characters
\usepackage[T1]{fontenc} % Output font encoding for international characters

\usepackage{subfiles}
\AtEveryBibitem{\clearfield{urlyear}\clearfield{urlmonth}}

\title{The role of antibody-mediated immunity in shaping the seasonality of respiratory viruses}

\author[a, b, $\dagger$]{Ruarai J Tobin}
\author[a, b]{James M McCaw}
\author[a, b, $\dagger$]{Freya M Shearer}

\affil[a]{\small School of Mathematics and Statistics, The University of Melbourne, Parkville 3010, Australia}

\affil[b]{\small Infectious Disease Dynamics Unit, Melbourne School of Population and Global Health, The University of Melbourne, Parkville 3010, Australia}

\affil[$\dagger$]{\small Corresponding authors: Ruarai J Tobin (ruarai.tobin@unimelb.edu.au) and Freya M Shearer (freya.shearer@unimelb.edu.au)}

\date{\vspace{-5ex}}
\begin{document}

\maketitle

\begin{abstract}
    In temperate regions, respiratory virus epidemics recur on a yearly basis, primarily during the winter season. This is believed to be induced by seasonal forcing, where the rate at which the virus can be transmitted varies cyclically across the course of each year. Seasonal epidemics can place substantial burden upon the healthcare system, with large numbers of infections and hospitalisations occurring across a short time period. However, the interactions between seasonal forcing and the factors necessary for epidemic resurgence --- such as waning immunity, antigenic variation or demography --- remain poorly understood. In this manuscript, we examine how the dynamics of antibody waning and antigenic variation can shape the seasonal recurrence of epidemics. We develop a novel susceptible-infectious-susceptible (\textit{SIS}) immuno-epidemiological model of respiratory virus spread, where the susceptible population is stratified by their antibody level against the currently circulating strain of the virus, with this decaying as both antibody waning and antigenic drift occur. In the absence of seasonal forcing, we demonstrate the existence of two Hopf bifurcations over the effective antibody decay rate, with associated periodic model solutions. When seasonal forcing is introduced, we identify complex interactions between the strength of forcing and the effective antibody decay rate, yielding myriad dynamics including multi-year periodicity, quasiperiodicity and chaos. The timing and magnitude of seasonal epidemics is highly sensitive to this interaction, with the distribution of infection timing (by time of year) varying substantially across the parameter space. Finally, we show that seasonal forcing can produce resonant damping resulting in a cumulative infection incidence that is less than would otherwise be observed.
\end{abstract}

\section{Introduction}

Respiratory viruses are a major cause of disease and mortality in human populations. It was estimated that in 2019, influenza led to between 290,000 and 640,000 deaths globally \parencite{Iuliano2018-tn} and in the same year, respiratory syncytial virus (RSV) led to between 84,000 and 125,000 deaths globally in children aged under five years alone \parencite{Li2022-ya}. Further, the COVID-19 pandemic has highlighted the risks that a newly emergent respiratory virus can pose, with the disease having caused at least five million deaths by the end of 2021 \parencite{Msemburi2023-sy}.

A defining characteristic of the transmission of respiratory viruses is that they often produce recurring epidemics (or waves) of infection and disease across human populations. In temperate regions, this recurrence can be highly regular --- for example, epidemics of the influenza viruses and respiratory syncytial virus are observed to occur every winter in temperate regions, with only rare exceptions \parencite{Altizer2006-yp, Moriyama2020-ez}. This is believed to arise as a result of seasonal forcing, where the rate at which a virus may be transmitted changes cyclically across the course of each year, typically increasing in winter and decreasing in summer. Such changes could be driven by seasonal conditions directly (e.g.\ changing temperature, humidity or ultraviolet radiation leading to changes in virus stability across the transmission process, \cite{Moriyama2020-ez, Nelson2004-ef}) or indirectly (e.g.\ more close contact occurring indoors during colder periods, \cite{London1973-hp, Moriyama2020-ez, Susswein2023-gm}). 

The arisal of a sustained epidemic wave of infection is only possible where a sufficient proportion of the population is susceptible to infection with the pathogen. Recovery from infection with a respiratory virus typically leads to the development of adaptive immunity, which protects individuals from re-infection with the same virus and results in a depletion of the susceptible population. As such, the continued recurrence of epidemic waves implies the existence of one or more mechanisms which may replenish the susceptible population \parencite{Keeling2008-pr}. A number of such mechanisms have been suggested for respiratory viruses, including: demographic processes (with the introduction of new susceptible individuals via births or migration); waning immunity (where the immunity developed following infection is lost over time); and antigenic variation (where the virus evolves to evade the immunity present in individuals previously infected). In this work, we focus on the role that waning immunity and antigenic variation may play in providing the conditions for epidemic recurrence.

The antibody response is a key component of human immunity to respiratory virus infection. This response involves the production of large numbers of antibodies, proteins that can inhibit or neutralise the pathogen's ability to replicate within the host \parencite{Murphy2019-kt}. For some respiratory viruses, high levels of pathogen-specific antibodies produced following infection have been associated with a reduced probability of an individual developing infection following exposure to the virus. For example, studies of influenza and SARS-CoV-2 have identified dose-dependent relationships between antibody levels and protection from (symptomatic) infection \parencite{Hobson1972-mm, Coudeville2010-re, Khoury2021-kg, Phillips2024-ug}. However, if the quantity of these antibodies were to decay (or that of the B cells which produce them), this immune protection would be temporary \parencite{Andraud2012-nr, Khoury2021-kg}. 

In addition to antibody decay, the effectiveness of an individual's antibody response may decay with time due to antigenic variation. Where this occurs, an individual's effective antibody level against the currently circulating virus (as measured via e.g.\ viral neutralisation assays) will be reduced \parencite{Webster1999-yk, Medina2011-it}. Such antigenic variation may occur as antigenic shift, where large changes in the antigen of the virus lead to sudden drops in the effectiveness of pre-existing antibodies, or as antigenic drift, where gradual changes in the virus accumulate and pre-existing antibodies become correspondingly less effective \parencite{Webster1999-nc, Kim2018-vx}. Such gradual antigenic drift is typical of many respiratory viruses, especially the influenza viruses \parencite{Webster1999-nc, Bedford2014-bd}.

In this paper, we construct a novel immuno-epidemiological model of respiratory virus transmission to study the interaction between antibody-mediated immunity and seasonal forcing. We begin in Section \ref{sec-model} by developing a susceptible-infectious-susceptible (\textit{SIS}) model of virus transmission, where the susceptible population is stratified by antibody level. This antibody level increases upon recovery from infection, decays exponentially over time due to the combined effects of antibody waning and antigenic drift, and modulates an individual's protection against re-infection. This combination of exponential decay and dose-dependent protection is informed by biological and epidemiological evidence and captures within a compartmental framework an approach to modelling antibody-mediated immunity that has become common within agent-based and statistical modelling frameworks \parencite{Hogan2023-ps, Muller2023-eu, Le2024-ku, Conway2024-wq, Yuan2025-le, Hao2025-io}. In Section \ref{sec-basic}, we show that this model can exhibit periodicity in the absence of seasonal forcing and identify two Hopf bifurcations across the antibody decay rate. This includes a subcritical Hopf bifurcation at a low effective antibody decay rate that yields multi-stability. In Section \ref{sec-seasonality}, we introduce seasonal forcing into the model and examine its effect upon epidemiological dynamics, performing a two-dimensional parameter sweep across the effective antibody decay rate and the strength of seasonal forcing. We apply a numerical classification approach to identify periodicity, quasiperiodicity and chaos, and use circular statistics to characterise seasonal behaviour. Our results show that forcing may interact with antibody-mediated immunity to produce complex dynamical regimes including quasiperiodicity and chaos, and identify that this interaction can lead to substantial changes in the distribution of infection timing by time of year. We also show how seasonal forcing can induce resonant dynamics that result in a cumulative infection incidence that is lesser than would arise in the absence of forcing. We end in Section \ref{sec-conclusions} with a brief conclusion.

\section{An immuno-epidemiological transmission model}\label{sec-model}

In this section, we construct an immuno-epidemiological model of respiratory virus transmission. We first classify our population of interest according to their epidemiological status, which may be either susceptible ($S$) or infectious ($I$). Individuals in the susceptible class have some effective antibody level $c$ against the currently circulating virus. We assume that this level decays exponentially with time due to a combination of antibody waning and antigenic drift, yielding an effective antibody decay rate $r$. For antibody waning, this fixed decay rate reflects a long-term exponential decay of antibodies due to depletion of plasma and memory B-cells that has been observed across a range of viruses (e.g.\ \cite{Andraud2012-nr, Movsisyan2022-yz, Deichmann2025-ti, Stocks2025-gq}). For simplicity, we have taken here that any contribution of antigenic drift to the effective antibody decay rate is fixed over time. This is consistent with the approximately linear increase in antigenic distance over time observed in, for example, the influenza viruses \parencite{Bedford2014-bd}. However, it does not capture the possibility of antigenic variation increasing during times of greater viral transmission \parencite{Bedford2015-zb}. In addition, this effective antibody level could only be directly observed in the scenario where no antigenic drift contributes to the effective antibody decay rate. Otherwise, the effective antibody level is a phenomenological quantity that measures the antibody response against an idealised `most recent' viral strain, which cannot be observed.

To capture this effective antibody level within a compartmental framework, we stratify the susceptible population across compartments $S_i$ where $0 \leq i \leq k$, each with a corresponding effective antibody level $c_i$:
\begin{equation*}
    c_i = 2^{a(i/k)},
\end{equation*}
where the parameter $a$ defines the maximal effective antibody level of $2^a$. Here, we describe these levels using powers of two in line with the common use of two-fold serial dilution assays for the measurement of the antibody response (e.g.\ \cite{Salk1944-aw, Killian2020-zq}). The exponential decay in effective antibody level is then captured using a method of lines approximation (Appendix A, \cite{Plant1986-jo, Angelov2023-gs}), with individuals in compartment $S_{i + 1}$ moving to compartment $S_i$ at a constant rate $\rho k$:
\begin{equation*}
\boxed{S_{i+1}} \xrightarrow[\mbox{\normalsize$\rho k$}]{} \boxed{S_i}
\end{equation*}
for $i \geq 0$ and $i < k$, where $\rho$ is specified in terms of the desired effective antibody decay rate $r$:
\begin{equation*}
    \rho = \frac{r}{a\log_e 2}.
\end{equation*}
These decay rates $r$ correspond to a half-life of the effective antibody level of $t_{1/2}=\log_e(2)/r$. For example, an effective antibody decay rate of $r = 0.02$ would correspond to an effective half-life of approximately 35 days, analogous to the rapid decay in effective antibody levels for a virus like SARS-CoV-2 with substantial antigenic drift \parencite{Lasrado2023-jq, Jeworowski2024-dc}. In comparison, a low effective antibody decay rate of $r = 0.005$ implies an effective half-life of 139 days, comparable to the relatively slow decay of antibodies observed for RSV, a virus for which antigenic drift is not a major contributor to re-infection \parencite{Falsey2006-zn}.

The parameter $k$ controls the resolution of the stratification (with the model having $k + 1$ strata over antibody level). For individuals starting at the same antibody level, varying $k$ changes the degree of between-individual variation in the effective antibody level as decay progresses (Appendix A). As such, $k$ may be specified to approximate the degree of dispersion expected in antibody decay (in a similar approach to \cite{Plant1986-jo}). As $k$ is increased towards infinity, this model approaches a partial differential equation where there is zero between-individual variation in the antibody decay process, which may not reflect biological reality. Given the uncertainties in specifying the degree of variation in decay rate, for the purposes of our study we consider an intermediate value of $k = 32$ and examine the effects of increasing $k$ in Appendix E.

In sum, we have a population described across $k + 2$ compartments:
\begin{equation*}
    \left\{S_0,\;S_1,\;\ldots,\;S_k,\;I\right\},
\end{equation*}
with the value of each compartment describing the fraction of the population which is within each state, such that the value of all compartments sums to one.

\subsection*{Immune protection and viral transmission}

We assume that individuals in the susceptible class have some level of immune protection against acquiring an infection with the currently circulating virus (conditional upon contact with an infected individual) which is dependent upon their effective antibody level. We take the protection $\omega_i$ for an individual in class $S_i$ to be a dose-response relationship over this effective antibody level $c_i$, specifically a Hill equation with a mid-point $c_\text{mid}$ and Hill coefficient $b$:
\begin{equation*}
\omega_i = \frac{c_i^b}{c_\text{mid}^b+c_i^b}.
\end{equation*}
This dose-response relationship has been previously used in both epidemiological studies of antibody-mediated immune protection \parencite{Halloran2009-py, Khoury2021-kg, Hogan2023-ps, Hao2025-io} and dynamic models of transmission \parencite{Zachreson2023-gp, Conway2024-wq}, though typically in the (mathematically equivalent) formulation of logistic regression over $\log c$. It is thought that this sigmoidal relationship arises due to saturation of viral antigen by antibody as the concentration of antibody increases \parencite{Parren1998-bm, Klasse2002-rn, Brandenberg2017-cz, Reeves2020-ot, Padmanabhan2022-kx}.

In addition to protecting against acquiring an infection, other potential effects of antibody-mediated immunity have been suggested, such as reducing the probability of developing severe disease or transmitting the infection onwards \parencite{Khoury2021-kg, Kim2023-bp, Wan2024-kz}, although there is a weaker consensus for these mechanisms strongly impacting transmission dynamics (for example, reductions in viral load are not consistently associated with pre-existing exposure, and the effect of a reduced viral load on infectiousness is not fully understood, \cite{Puhach2023-gj, Lingas2024-ci, Singanayagam2022-zo, Jones2021-za, Zhou2023-al}). In Appendix F, we consider a version of the model where antibody-mediated immunity also reduces the duration of infectiousness, where we find that epidemiological-scale dynamics are dominated by whichever of the two immune mechanisms arises at a lower effective antibody level (i.e.\ a lower mid-point in the antibody-effect curve). Further, we find that there is little qualitative difference in the resultant dynamics between the two scenarios that could not be captured by a reparametrisation of the model with only antibody-mediated protection against infection. As such, we consider only the effects of antibody-mediated immunity in reducing the probability of acquiring an infection for the remainder of this work.

In our model, contact between individuals occurs according to mass-action effect. A susceptible individual makes infectious contacts with infected individuals at a rate given by the product of the proportion of infectious individuals $I$ and the infectiousness parameter $\beta$. Individuals in each susceptible strata $S_i$ are then infected according to this rate of exposure, reduced by their level of antibody-mediated protection $\omega_i$:
\begin{equation*}
\boxed{S_i} \xrightarrow[\mbox{\normalsize$(1 - \omega_i)\beta I$}]{} \boxed{I} 
\end{equation*}

\subsection*{Recovery from infection and post-infection immunity}

Infectious individuals are assumed to recover from infection at a fixed rate $\gamma$ (such that the mean duration of infectiousness is $1/\gamma$). Upon recovery from infection, individuals transition to a susceptible strata $S_i$ according to the probability $\text{P}(I \rightarrow S_i)$:
\begin{equation*}
\boxed{I} \xrightarrow[\mbox{\normalsize$\gamma \text{P}(I \rightarrow S_i)$}]{} \boxed{S_i} 
\end{equation*}

We specify $\text{P}(I \rightarrow S_i)$ in terms of an assumed distribution of effective antibody level following recovery from infection. Specifically, we take that the log-2-transformed value of the effective antibody level follows a Normal distribution with mean $\mu$ and standard deviation $\sigma$. This distribution then defines the probability of transitioning into each discrete susceptible strata, noting we have an additional constraint that antibody level must fall within the range $[2^0, 2^a]$:
\begin{equation*}
    \text{P}(I \rightarrow S_i) = \begin{cases}
        \text{P}(a < X), \quad &i = k,\\
        \text{P}\left(a\frac{i}{k} < X \leq a\frac{i+1}{k}\right), \quad &i > 0 \text{ and } i < k,\\
        \text{P}\left(X \leq a\frac{1}{k}\right), \quad &i = 0.\\
    \end{cases}
\end{equation*}
The use of a Normal distribution here is consistent with epidemiological observations of the antibody response following exposure to respiratory viruses such as SARS-CoV-2 (e.g.\ \cite{Khoury2021-kg, Yamayoshi2021-py}) and influenza (e.g.\ \cite{Zhao2017-rq}). This Normal distribution likely arises due to heterogeneity in the humoral immune response (antigen dose, rate of antibody growth, etc.) between infection events \parencite{Padmanabhan2022-kx, Stocks2025-gq}.

The model of immunity we describe here captures only the first-order dynamics of antibody rise and decay within an individual. To simplify our analysis, we do not consider the dynamics of other components that make up the human adaptive immune response, such as plasma cells (which produce antibodies), memory B-cells (which recall the antibody response upon re-infection) or T-cells (which may provide protection independent of the antibody response). If these other components behaved in a similar manner to the antibody response but provided protection across greater durations, their presence could be seen as equivalent to a reduced decay rate $r$. However, more complex behaviour such as interactions between the different components across repeat infections cannot be captured by our one-dimensional model. Further, we do not capture any process of demographic change in this model, which has been previously identified as an important factor in similar modelling studies \parencite{Dafilis2012-yo}. If this process were to act via the removal of individuals at random and the addition of immune-naive individuals, a greater rate of demographic turnover would be approximately equivalent to an increase in the effective antibody decay rate. 

This model captures key components of antibody-mediated immunity for respiratory viruses, including an exponential decay of antibody level, a dose-response relationship between the antibody level and immune protection, and a post-infection distribution of antibody levels. This reflects assumptions that have become common practice in modelling respiratory virus immunity and transmission, particularly in application to SARS-CoV-2 due to the wealth of data that was accumulated during the pandemic. For example, a number of agent-based models were developed to understand potential future immunity scenarios for SARS-CoV-2 included these relationships to define immune protection across populations (e.g.\ \cite{Muller2023-eu, Le2024-ku, Conway2024-wq, Yuan2025-le}). In addition, these mechanisms have been captured within statistical models to predict immune protection in different population groups (e.g.\ \cite{Hogan2023-ps, Hao2025-io}). In capturing these mechanisms within a deterministic compartmental framework, we are able to explore the dynamical implications of these mechanisms on epidemiological outcomes across long time-scales.

\subsection*{Model structure}

\begin{figure}[H]
    \centering
    \makebox[\textwidth][c]{\includegraphics[width=12cm]{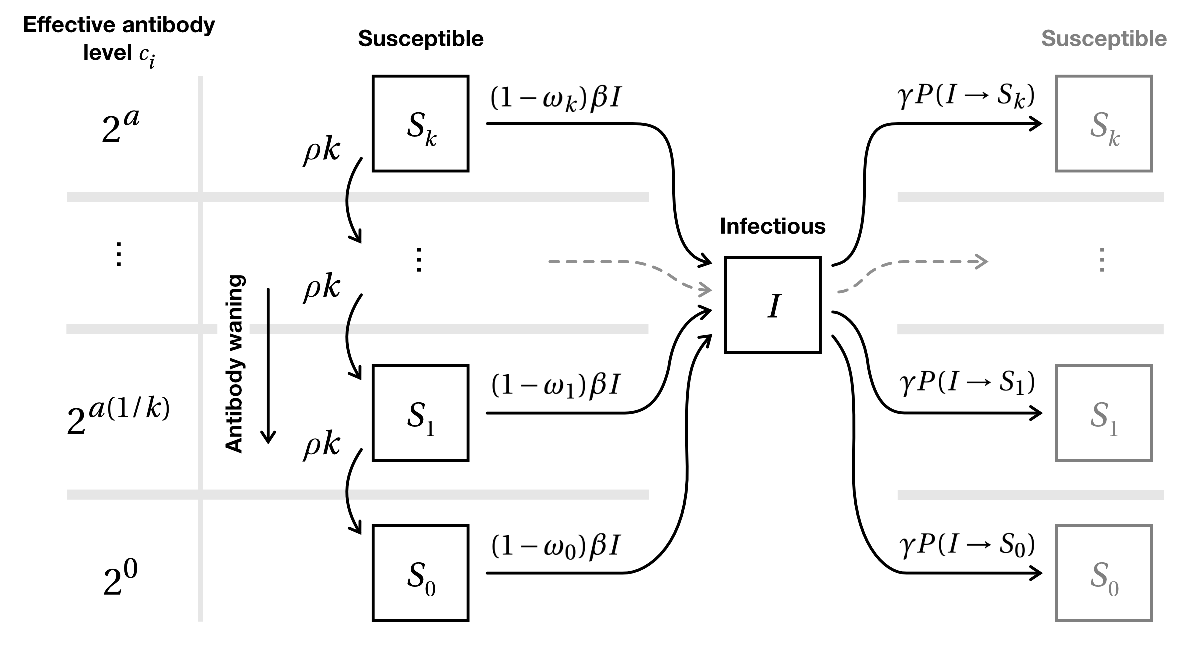}}
     \caption{Compartmental flow diagram of our immuno-epidemiological model of respiratory virus transmission. The susceptible class $S_i$ is stratified according to their effective antibody level $c_i$ (which may range from $2^0$ up to $2^a$), with this defining $\omega_i$, the level of immune protection against infection given contact with an infected individual.}
    \label{compartmental-model}
\end{figure}

Combining the above model transitions, we have a deterministic dynamical system defined by $k + 2$ ordinary differential equations:
\begin{align}
    \dv{S_i}{t} &= \begin{cases}
        - \beta (1-\omega_i) S_i I + \gamma \text{P}(I \rightarrow S_i) I \hphantom{+ \rho k S_{i + 1}}\;\; - \rho k S_i, & i = k,\\[0.7em]
        - \beta (1-\omega_i) S_i I + \gamma \text{P}(I \rightarrow S_i) I + \rho k S_{i + 1} - \rho k S_i,& 0 < i < k,\\[0.7em] 
        - \beta (1-\omega_i) S_i I + \gamma \text{P}(I \rightarrow S_i) I + \rho k S_{i + 1},& i = 0 , 
    \end{cases}\label{eq-model-1}\\[1.0em]
    \text{and}\nonumber\\
    \dv{I}{t} &= \beta I \sum_{i=0}^k (1 - \omega_i)S_i - \gamma I,\label{eq-model-2}
\end{align}
which we illustrate in Figure \ref{compartmental-model}.

The model we describe is similar in nature to the general recovery-stratified model described by \textcite{El-Khalifi2023-cd}, but differs in our allowance for variability in the post-infection antibody level and their use of different functions to describe immune protection across the different strata.

\subsection*{Numerical methodology}

All results were produced using the parameter values listed in Table \ref{table-params} (except where the effective antibody decay rate $r$ was otherwise varied). These values correspond to a moderately infectious respiratory virus with a basic reproduction number ($\text{R}_0$) of 1.5 and a generation interval of four days. At baseline, antibody-mediated protection was specified such that at least 50\% protection from re-infection was maintained following recovery from infection for a median period of approximately 139 days. We set the initial conditions of the system such that a small proportion of the population was infectious ($I = 0.001$), while the remaining population was in the susceptible class at the minimum effective antibody level ($S_0 = 0.999$).

Where we calculated infection incidence (i.e.\ the number of new infections across some period), we included in our numerical implementation an additional cumulative infections compartment $\text{C}_I$, defined as:
\begin{equation*}
    \dv{\text{C}_I}{t} = \beta I \sum_{j=0}^k (1 - \omega_j)S_j,
\end{equation*}
where daily infection incidence for day $t$ is then calculated as $\text{inc}(t) = \text{C}_I(t + 1) - \text{C}_I(t)$. 

We identified the (stable or unstable) fixed points of the system by first reducing the systems dimensionality by taking $I = 1 - \sum_{i=0}^kS_i$ and setting the left hand side of this system to zero and solving for $\{S_0, S_1, \ldots, S_k\}$ under the constraint that each compartment value was positive. A fixed point solution was defined as stable where all eigenvalues of the reduced linearised system had real components less than zero.

We assessed numerical solutions for chaos using the Melbourne 0-1 test across the numerically integrated infection prevalence time-series $I(t)$ following 1,000 years of burn-in \parencite{Gottwald2016-jg}. To avoid known issues with the test regarding the oversampling of continuous-time dynamical systems, we performed the test over infection prevalence sampled once every 80 days. This downsampling rate was identified visually to achieve good classification performance over a subset of the results (Appendix Figure \ref{figure-supp-chaos-downsample}). To distinguish between periodic, quasiperiodic and chaotic solutions (and identify the duration of the periodic solutions), we used a classification algorithm which we describe in Appendix C.

All numerical analysis was performed in Julia \parencite{Bezanson2017-ym}. Numerical integration of the dynamical system was performed using the 5th order Rosenbrock-Wanner method \textit{Rodas5P} as implemented in the DifferentialEquations.jl package \parencite{Rackauckas2017-gy, Steinebach2023-ka}. NonlinearSolve.jl was used for identifying the fixed points of the system \parencite{Rackauckas2023-rn}, and DynamicalSystems.jl was used for the Melbourne 0-1 test for chaos \parencite{Datseris2018-kp, Datseris2022-ov}. Normal forms of the Hopf bifurcations identified in the system without forcing were calculated using BifurcationKit.jl \parencite{Veltz2020-ak}. Visualisation of results was performed in the R statistical computing environment using the ggplot2 package \parencite{R-Core2013-ut, Wickham2011-gk}.

The code used to produce the results is available on GitHub (\url{https://github.com/ruarai/ode_immunity}) and has been archived on OSF (\url{osf.io/us9t8}). 

\begin{table}[H]
\footnotesize
\begin{tabular*}{\textwidth}{@{}l|l@{}}
\toprule
Parameter                                                       & Value \\ \midrule
Maximum strata index                                            & $k = 32$ \\
Maximal effective antibody level                                & $2^a = 2^{8.0}$ \\
Effective antibody decay rate (where otherwise unchanged)       & $r = 0.015 \text{ days}^{-1}$ \\
Recovery rate                                                   & $\gamma = 0.25 \text{ days}^{-1}$ \\ 
Transmission rate                                               & $\beta = 0.375 \text{ days}^{-1}$ \\ % Because compartments are unitless
Post-infection antibody level                                   & $X \sim \text{N}(6, 0.5)$ \\
Midpoint of the protection function                             & $c_\text{mid} = 2^3$ \\
Hill coefficient of the protection function                     & $b = 8$\\ \midrule
\end{tabular*}
\caption{Model parameter values used in the results.}
    \label{table-params}
\end{table}

\section{Model dynamics}\label{sec-basic}

\begin{figure}[H]
    \centering
    \makebox[\textwidth][c]{\includegraphics[width=13cm]{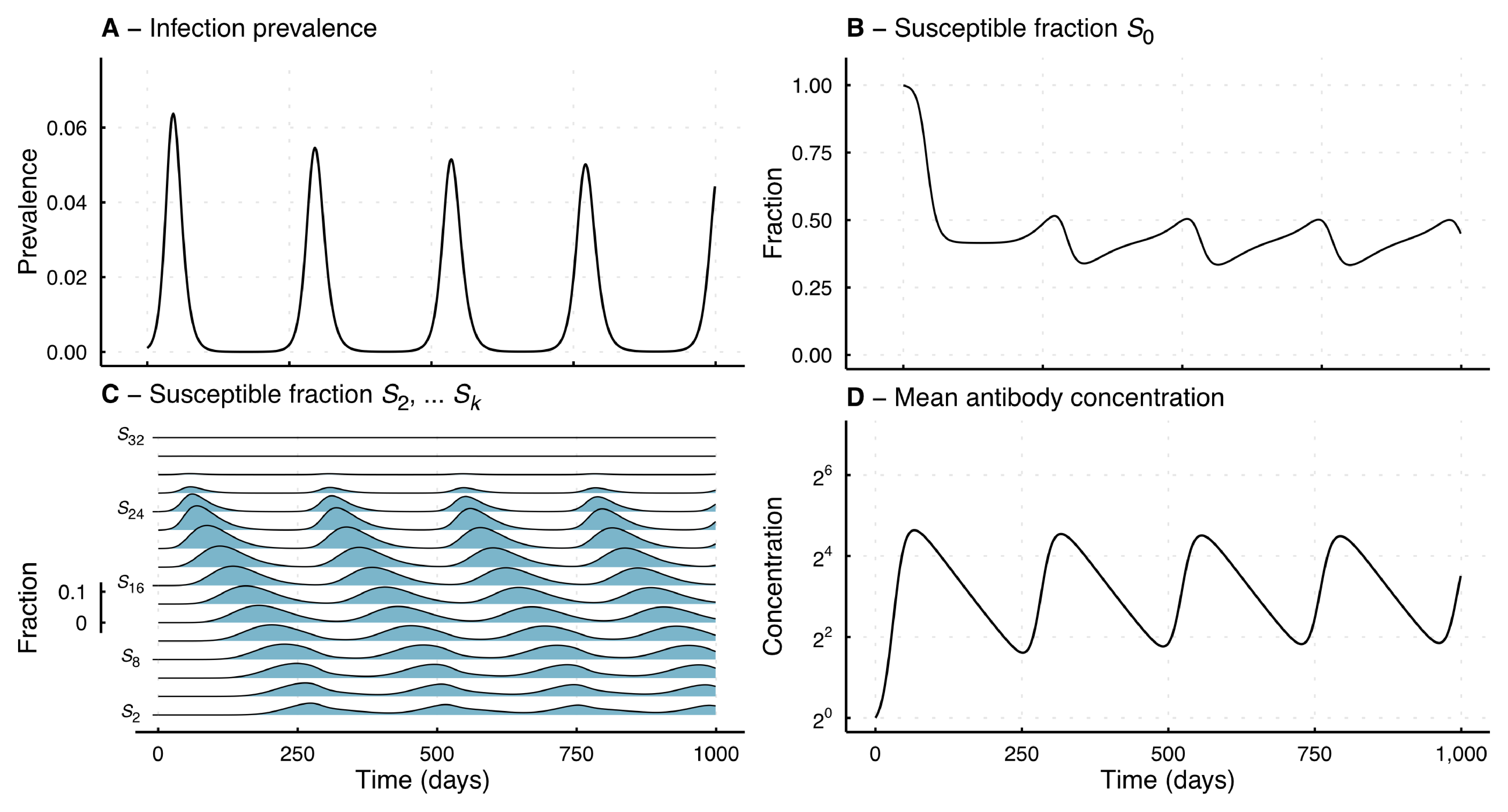}}
     \caption{Characteristic dynamics of the immunity-structured model of respiratory virus transmission. Model parameters are as in Table \ref{table-params}. \textbf{A}: Prevalence (i.e.\ fraction) of infectious individuals $I$ over time. \textbf{B}: Fraction of susceptible individuals in the minimum antibody strata $S_0$. \textbf{C}: Fraction of susceptible individuals in the strata $S_2$ through to $S_k$ (odd strata not plotted for visual clarity), with each strata plotted as an individual ribbon. \textbf{D}: The mean effective antibody level across the susceptible population, calculated as an arithmetic mean (i.e.\ not a geometric mean). }
    \label{figure-basic}
\end{figure}

We illustrate the characteristic dynamics produced by the transmission model in Figure \ref{figure-basic}, with parameters as specified in Table \ref{table-params}. Across the first 125 days, we observe an SIR-like wave of infection (Figure \ref{figure-basic}A), with over half of the population initially in $S_0$ being infected (Figure \ref{figure-basic}B). Following recovery from infection, the fraction of the population within the higher-index susceptible classes increases (Figure \ref{figure-basic}C), in line with the assumed distribution of post-infection effective antibody levels. The population then experiences antibody waning, with the fraction within the lower-index susceptible classes increasing. Once there is a sufficiently large density of susceptible individuals in lower-index (and lower-protection) classes, a subsequent wave of infection occurs. This cycle repeats indefinitely undamped. Such oscillatory behaviour is typical of immuno-epidemiological models where the population is structured by immunity \parencite{Heffernan2009-mz, Bohm2016-re, Barbarossa2018-ha}. We illustrate the complete distribution of individuals across the susceptible strata over time in Appendix Figure \ref{figure-supp-basic}. The arithmetic mean of effective antibody level also varies substantially with time (Figure \ref{figure-basic}D). Starting at a mean of $2^0$ (with all susceptible individuals in $S_0$), the initial wave of infection leads to a peak mean effective antibody level of approximately $2^{4.5}$, and between waves of infection the level reaches a minimum of around $2^2$. 

\begin{figure}[H]
    \centering
    \makebox[\textwidth][c]{\includegraphics[width=14cm]{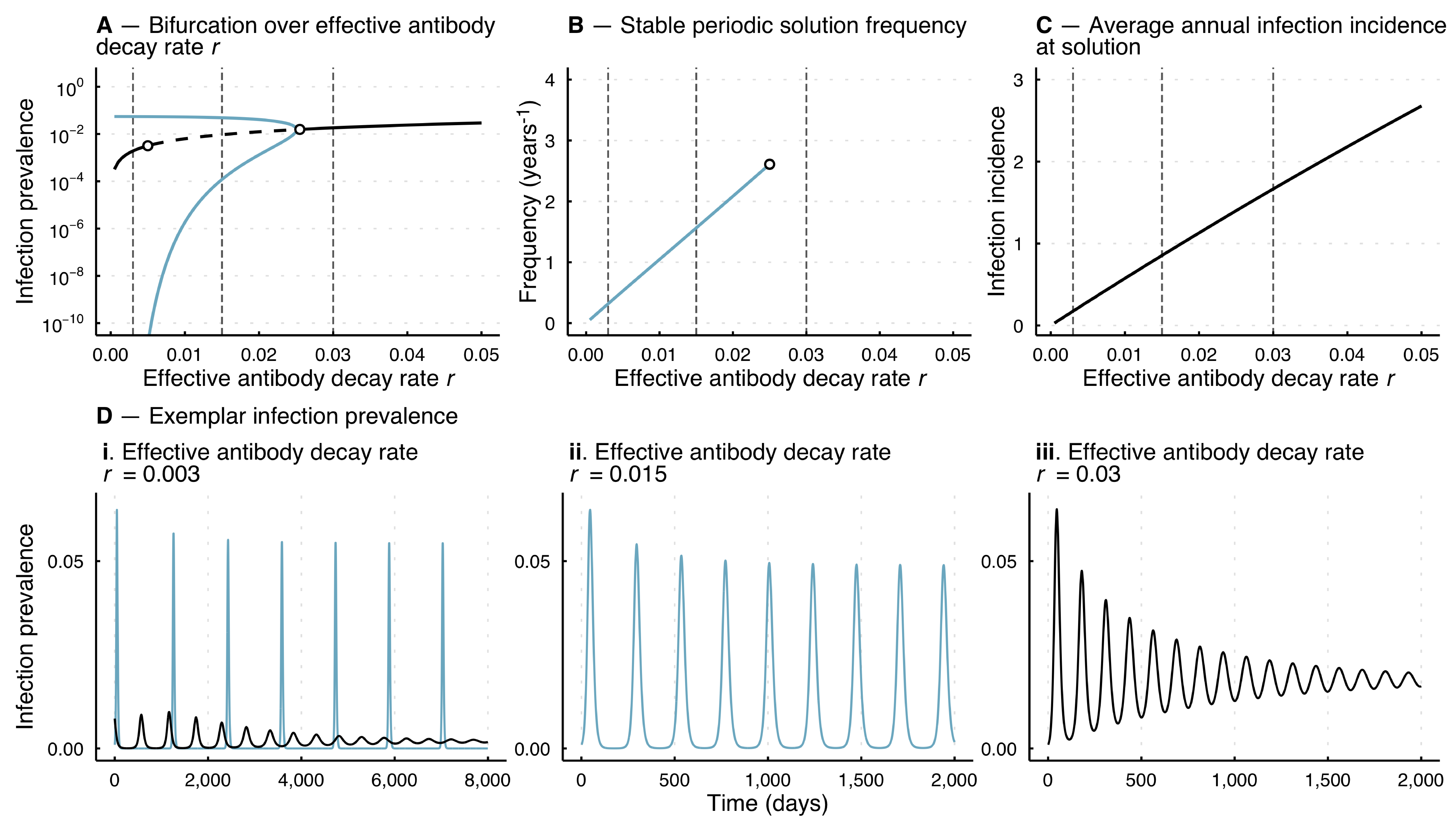}}
     \caption{Dynamics of the immuno-epidemiological model for varying values of effective antibody decay rate $r$ (all other model parameters are as in Table \ref{table-params}). The system was evaluated across a period of 100 years (36,500 days), following 100 years of burn-in. \textbf{A}: Bifurcation diagram over varying effective antibody decay rate $r$, with lines depicting infection prevalence at the model solutions (noting the log scale on the y-axis). For the periodic solutions, the light blue lines correspond to the maximal or minimal infection prevalence across each solution. For low values of the decay rate (e.g.\ less than $r = 0.01\;\text{days}^{-1}$), the minimal infection prevalence rapidly approaches values where stochastic effects would be dominant in realistic settings and our deterministic results must be interpreted with care. The separatrix between the stable limit cycle and the stable fixed point for values of $r$ less than approximately $0.005\;\text{days}^{-1}$ is not depicted as it could not be identified numerically. \textbf{B}: The frequency (in years$^{-1}$) of the periodic model solutions for varying effective antibody decay rate $r$. \textbf{C}: The average annual infection incidence at the model solution (across both fixed point and stable periodic solutions) across effective antibody decay rate $r$, calculated as the mean daily infection incidence multiplied by 365. \textbf{D}: Exemplar dynamics in infection prevalence across the three dynamical regimes, corresponding to the vertical dashed lines in \textbf{A}-\textbf{C}. Those which tend towards a stable periodic solution are illustrated in blue while those which tend towards a stable fixed point are black. The solution for $r = 0.003\;\text{days}^{-1}$ (panel \textbf{i}) which tends towards the stable fixed point was initialised by sampling a random point in state space near the fixed point. Note that the x-axis extent differs for panel \textbf{i}.}
    \label{figure-bifurcation}
\end{figure}

We now examine the effect of varying the effective antibody decay rate $r$. For the range of effective antibody decay rates explored ($0<r\leq 0.05 \;\text{days}^{-1}$), we observe a variety of dynamical outcomes (Figure \ref{figure-bifurcation}A). For high values of decay rate (e.g.\ $r = 0.03$), infection prevalence tends towards a stable fixed point (e.g.\ Figure \ref{figure-bifurcation}D panel \textbf{iii}). As the decay rate is reduced, a Hopf bifurcation occurs (at $r \approx 0.025\;\text{days}^{-1}$), with the fixed point solution becoming unstable and a stable periodic solution emerging (e.g.\ Figure \ref{figure-bifurcation}D panel \textbf{ii}). This Hopf bifurcation is evident in a pair of conjugate eigenvalues of the linearised system crossing the imaginary axis (Appendix Figure \ref{figure-supp-eigenvalues}), and a first Lyapunov coefficient of $\alpha \approx -0.356$ indicates that the bifurcation is supercritical. A similar Hopf bifurcation is identified in \textcite{Hethcote1981-gj}, where periodic solutions arise in an SIRS delay-differential equation model only where individuals are ``significantly delayed'' in the temporarily immune class. It is likely that the periodicity we observe in our model arises via a similar delay process induced by antibody waning.

As the decay rate is reduced further, a second Hopf bifurcation occurs at $r \approx 0.005\;\text{days}^{-1}$, with the fixed point solution again becoming stable and the pair of conjugate eigenvalues returning to the left hand side of the complex plane (Appendix Figure \ref{figure-supp-eigenvalues}). Following the second Hopf bifurcation, the periodic solution produced by the first Hopf bifurcation remains stable, implying the existence of an unstable separatrix that divides the two basins of attraction between the stable periodic solution and the stable fixed point solution. The first Lyapunov coefficient at this bifurcation is $\alpha \approx 0.357$, indicating it is subcritical. We illustrate two trajectories in this regime ($r = 0.003\;\text{days}^{-1}$, Figure \ref{figure-bifurcation}D panel \textbf{i}), one which tends towards the stable periodic solution and one which tends towards the stable fixed point solution.

We identify that the minimal infection prevalence across the stable periodic solution declines rapidly as the effective antibody decay rate $r$ is decreased (Figure \ref{figure-bifurcation}A). At the extremely low minimal infection prevalences that we observe (e.g.\ $I < 10^{-7}$), it would be expected that, in any realistic setting, the dynamics of transmission around this minimum would be highly stochastic. For example, in a population size of $10^8$, this would correspond to less than ten concurrent infections on average at the trough between each wave, implying a high probability of pathogen extinction. In this scenario, the sustained transmission of a virus would only be possible where the antibody-mediated immunity across the population could otherwise be avoided (e.g.\ through demographic change). However, the existence of a second Hopf bifurcation implies that a virus which induces a particularly slow decaying antibody response (less than $r \approx 0.005\;\text{days}^{-1}$) could evade extinction at the stable fixed point, where infection prevalence is relatively high. This numerical analysis does not preclude the possibility of additional attractors existing outside of the immediate neighbourhood of the fixed point in the global state space or as other parameters are varied.

We find that both the frequency of the stable periodic solution and the average annual infection incidence increase approximately linearly with effective antibody decay rate $r$ (Figure \ref{figure-bifurcation}B, C). This average annual infection incidence is calculated as the average daily infection incidence in the period following 100,000 days of burn-in, multiplied by 365. Notably, where both a fixed point solution and a stable periodic solution are present ($r$ less than approximately $0.025\;\text{days}^{-1}$), this average annual infection incidence is equal between the two solutions.

\subsection*{Stratification of the infectious class and antibody boosting}

In Appendix B, we provide a specification of the model where the infectious class is also stratified by effective antibody level. In this version of the model, a susceptible individual in $S_i$ who is infected enters the corresponding infectious compartment $I_i$. Infectious individuals then recover and return to a susceptible compartment $S_j$ according to a probability distribution $\text{P}(I_i \rightarrow S_j)$. This allows us to specify a dependence between the pre-infection and post-infection effective antibody level $c_i$ and $c_j$.

We use this model to examine the effect of `antibody boosting', where an individual's post-infection effective antibody level is an increasing function of their pre-infection effective antibody level \parencite{Barbarossa2018-ha, Diekmann2018-ev}. Our results (Appendix Figure \ref{figure-boosting-results}) show that the inclusion of antibody boosting may quantitatively alter the dynamics produced by the model, changing both the peak infection prevalence and peak mean effective antibody level. However, the overall qualitative nature of the dynamics are unchanged. As in Figure \ref{figure-bifurcation}, we identify two Hopf bifurcations across the effective antibody decay rate, an (approximate) linear relationship between the decay rate and the periodic solution frequency, and an (approximate) linear relationship between the decay rate and the average annual infection incidence. Given the lack of qualitative differences in equilibrium dynamics, in the following sections we consider the model without stratification of the infectious class.

\section{Waning, seasonal forcing and periodicity}\label{sec-seasonality}

In the above section, we show how our model can produce oscillatory behaviour in the absence of any external forcing. As previously discussed, such periodic model dynamics are typical of immuno-epidemiological models (e.g.\ \cite{Heffernan2009-mz, Bohm2016-re, Barbarossa2018-ha}). However, where \textcite{Heffernan2009-mz} and similar works have primarily focussed on aspects such as how epidemiological dynamics may be affected by different assumptions regarding within-host immunity, our focus here is on how these immune dynamics may interact with the external seasonal forcing that is typical of models of seasonal respiratory virus transmission. As such, in this section we extend the model to include a seasonally-varying transmission rate and examine how this affects the long-term epidemiological dynamics. To do this, we consider a time-varying infectiousness term:
\begin{equation*}
    \beta(t) = \beta_0 \left(1 + \eta \cos(\frac{2 \pi}{365} t)\right),
\end{equation*}
such that infectiousness varies sinusoidally over a period of 365 days, with a maximal value of $\beta_0(1+\eta)$ at the start of each year, and a minimal value of $\beta_0(1-\eta)$ mid-year (corresponding to seasonal effects in a temperate region in the Northern Hemisphere). In the following, we take $\beta_0 = 0.375$, such that the average value of $\beta(t)$ matches the constant $\beta$ used in the previous section.

\begin{figure}[H]
    \centering
    \makebox[\textwidth][c]{\includegraphics[width=13cm]{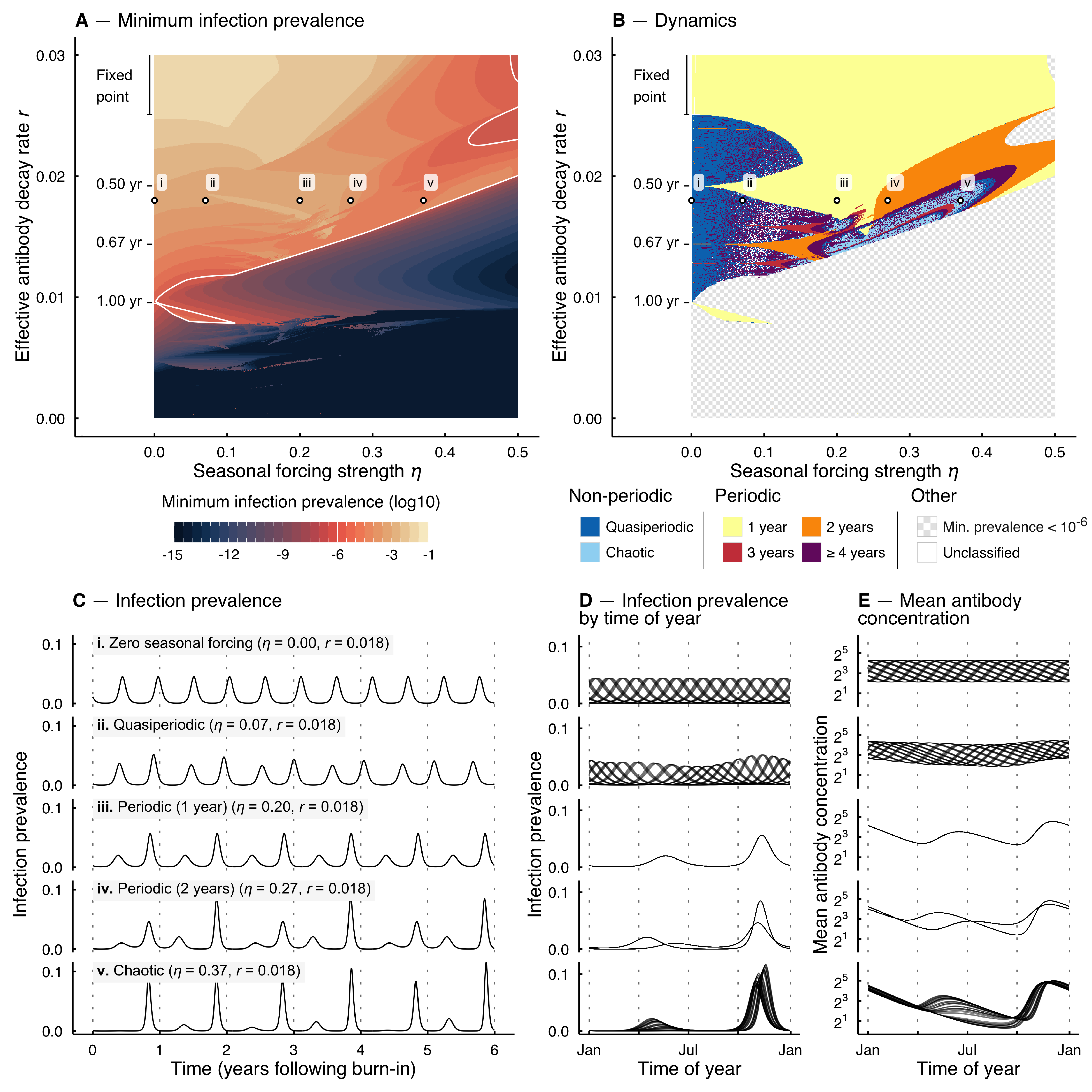}}
     \caption{Model dynamics across varied seasonal forcing strength $\eta$ and effective antibody decay rate $r$. Model parameters are otherwise as in Table \ref{table-params}. The system was evaluated across a period of 250 years, following 1,000 years of burn-in. \textbf{A}: The qualitative dynamics yielded following burn-in for each parameter pair. Our approach to classifying the dynamics of the numerically integrated model solutions is described in Appendix C. Classifications for results which had a low minimum infection prevalence (less than $10^{-6}$) were masked. The complete classifications including those masked are displayed in Appendix Figure \ref{figure-supp-seasonality-full}. \textbf{B}: The minimum infection prevalence across the 250 years following burn-in for each parameter pair. For realistic settings, results with a low minimum (e.g.\ less than $10^{-7}$) would likely be dominated by stochastic effects. \textbf{C}: Example trajectories of infection prevalence across six years following burn-in. \textbf{D}, \textbf{E}: Example trajectories of infection prevalence and mean effective antibody level respectively, indexed by time of year. Trajectories from 60 consecutive years following burn-in are displayed.}
    \label{figure-grid-seasonality}
\end{figure}

Across values of the effective antibody decay rate $r$ ranging from zero (no antibody decay) to 0.1 (high decay rate) and values of the strength of seasonal forcing $\eta$ ranging from zero (no seasonal forcing) to 0.5 (very high seasonal forcing), a diverse range of long-term dynamics are identified (Figure \ref{figure-grid-seasonality}A, \ref{figure-grid-seasonality}B). This includes periodic solutions (with periods of one or more years), quasiperiodic solutions and chaotic dynamics, where we have defined chaos as a positive result on the Melbourne 0-1 test (Appendix C). The minimum infection prevalence we observe (across the 250 years following burn-in) varies substantially across this parameter space (Figure \ref{figure-grid-seasonality}A). The minimum infection prevalence has the highest value where seasonal forcing is absent and the effective antibody decay rate is greatest (i.e.\ $\eta = 0$ and $r = 0.03 \;\text{days}^{-1}$), and tends to reduce as either the strength of seasonal forcing $\eta$ is increased or the effective antibody decay rate $r$ is decreased. Generally, results where seasonal forcing is present have substantially lower values of minimum infection prevalences than those without forcing (Appendix Figure \ref{figure-supp-seasonality-extinction}). As in the case without seasonal forcing, the results which reach particularly low infection prevalence would, in realistic settings, have greater probabilities of stochastic extinction occurring. As such, we mask the results of parameter combinations that produce an infection prevalence less than $10^{-6}$ in Figure \ref{figure-grid-seasonality}B, with the same masking performed for results presented later in this work. The full results of Figure \ref{figure-grid-seasonality}B without masking are presented in Appendix Figure \ref{figure-supp-seasonality-full}.

Across this parameter space, the occurrence of periodic solutions is dependent upon the natural period of the system (i.e.\ the period in the absence of seasonal forcing, $\eta = 0$). In Figure \ref{figure-grid-seasonality}A and \ref{figure-grid-seasonality}B, we annotate the y-axis with this natural period for a subset of rational multiples of the one-year period of seasonal forcing. As the seasonal forcing strength $\eta$ is increased from zero, periodic solutions appear in triangular clusters around these rational multiples, with the period in years of the solution equal to the numerator of the rational multiple. These regions are known as `Arnold tongues' \parencite{Datseris2022-ov}. Each rational multiple of the natural period has a corresponding Arnold tongue, however, only a small number are visible in our visualisation as they typically become vanishingly small as the numerator or denominator increases \parencite{Datseris2022-ov}. The presence of these Arnold tongues implies that --- for certain effective decay rates --- periodicity in infection dynamics could be induced by mechanisms which only have slight effects upon the rate at which the virus may be transmitted across each year. For example, even a 10\% difference in the rate at which individuals make contact between the summer and winter periods of the year (i.e.\ $\eta \approx 0.05$) leads to periodicity across a wide range of effective antibody decay rates. This identification of high sensitivity of periodicity to seasonal forcing aligns with similar findings made using a stochastic SIRS model with an exponentially distributed period of immunity \parencite{Dushoff2004-wv}.

In the space bounded between these Arnold tongues, and for effective antibody decay rates less than $0.025\;\text{days}^{-1}$ (where, in the absence of seasonal forcing, stable periodic solutions are present), quasiperiodic solutions can be identified (e.g.\ Figure \ref{figure-grid-seasonality}C, panel \textbf{ii}). For these quasiperiodic solutions, the duration of time between each wave of infection is relatively predictable, but the timing of these waves shifts relative to the calendar year each year, such that epidemics may peak at any time of year in the long-run (Appendix Figure \ref{figure-supp-peak-timing}). Hypothetically, a newly emergent respiratory virus could exhibit such quasiperiodic dynamics in the long-term if it were to become endemic within the human population. This could present a challenge to public health planning, as any strategies predicated upon the regular seasonal recurrence of an epidemic would be undermined. Public health activities such as pathogen surveillance, vaccination deployment, and the mobilisation of healthcare workers --- rather than beginning and ending on an approximately regular calendar schedule --- would need to be adjusted each year, and would often fall outside the winter epidemic season. However, epidemics that occur outside of the usual wintertime period are less likely to be co-circulating with other viruses, and hence the total clinical load from respiratory infections might be lower compared to winter months.

As the seasonal forcing strength $\eta$ increases, the growing Arnold tongues collide with one another, and regions of chaotic dynamics are evident. Period-doubling bifurcations can be identified at the border of some of these chaotic regions (Appendix Figure \ref{figure-supp-chaos-downsample}). Notably, the presence of chaos does not necessarily imply that the long-term qualitative dynamics of infection are difficult to anticipate or manage. For example, the trajectory exhibiting chaotic dynamics that we present (Figure \ref{figure-grid-seasonality}D, panel \textbf{v}) exhibits relatively little variation in the timing and magnitude of the primary (larger) wave of infection each year. Rather, the effects of chaos are more pronounced in the secondary (smaller) wave of infection, which has a size that varies substantially year to year.

We find that the rate at which peaks in infection incidence occur across the period following burn-in may vary widely as the strength of seasonal forcing and the effective antibody decay rate change (Appendix Figure \ref{figure-supp-seasonality}A, B). Recent work by \textcite{Rubin2025-du} has investigated the possibility that the `sub-annual' recurrence of SARS-CoV-2 --- in which multiple waves of infection occur each year --- which has been observed in many regions could be driven by the dynamics of immunity. We identify similar results, with lower effective antibody decay rates yielding multiple peaks in infection incidence per year following burn-in (Appendix Figure \ref{figure-supp-seasonality}B).

In Appendix E we consider a sensitivity analysis of the results presented in Figure \ref{figure-grid-seasonality} across a range of model scenarios. In Appendix Figure \ref{figure-supp-seasonality-sensitivity}, we vary parameters that define the dynamics of antibody-mediated immunity. We find that increasing dispersion in the decay process (e.g.\ increasing the variance of the post-infection antibody distribution or lowering the Hill coefficient of the protection curve) tends to lead to replacement of large regions of quasiperiodicity and chaos with simple period-1 dynamics. In contrast, the bifurcation structure is generally preserved where dispersion is reduced. In particular, as $k$ is increased (increasing the number of antibody strata and thus reducing dispersion in the decay process as described in Appendix A), the bifurcation structure is preserved. In addition, in Appendix Figure \ref{figure-supp-seasonality-sensitivity-strat}, we show that the inclusion of antibody boosting has relatively little impact on results. Furthermore, in Appendix G we apply the same numerical bifurcation analysis methodology to a seasonally-forced \textit{SIRS} model (Appendix Figure \ref{figure-supp-seasonality-SIRS}). We find a similar simplification of the bifurcation structure as in the sensitivity analyses where dispersion across the dynamics of antibody-mediated immunity is increased (i.e.\ Appendix Figure \ref{figure-supp-seasonality-sensitivity}). In particular, Arnold tongues are less apparent. These results demonstrate that the \textit{SIRS} model may be a suitable approximation for our model in the case where dispersion in the dynamics of antibody-mediated immunity is substantial, but would fail to capture the more complex dynamical behaviours we observe in Figure \ref{figure-grid-seasonality} where lower dispersion is assumed.

\subsection*{Annual timing of infection incidence}

\begin{figure}[H]
    \centering
    \makebox[\textwidth][c]{\includegraphics[width=13cm]{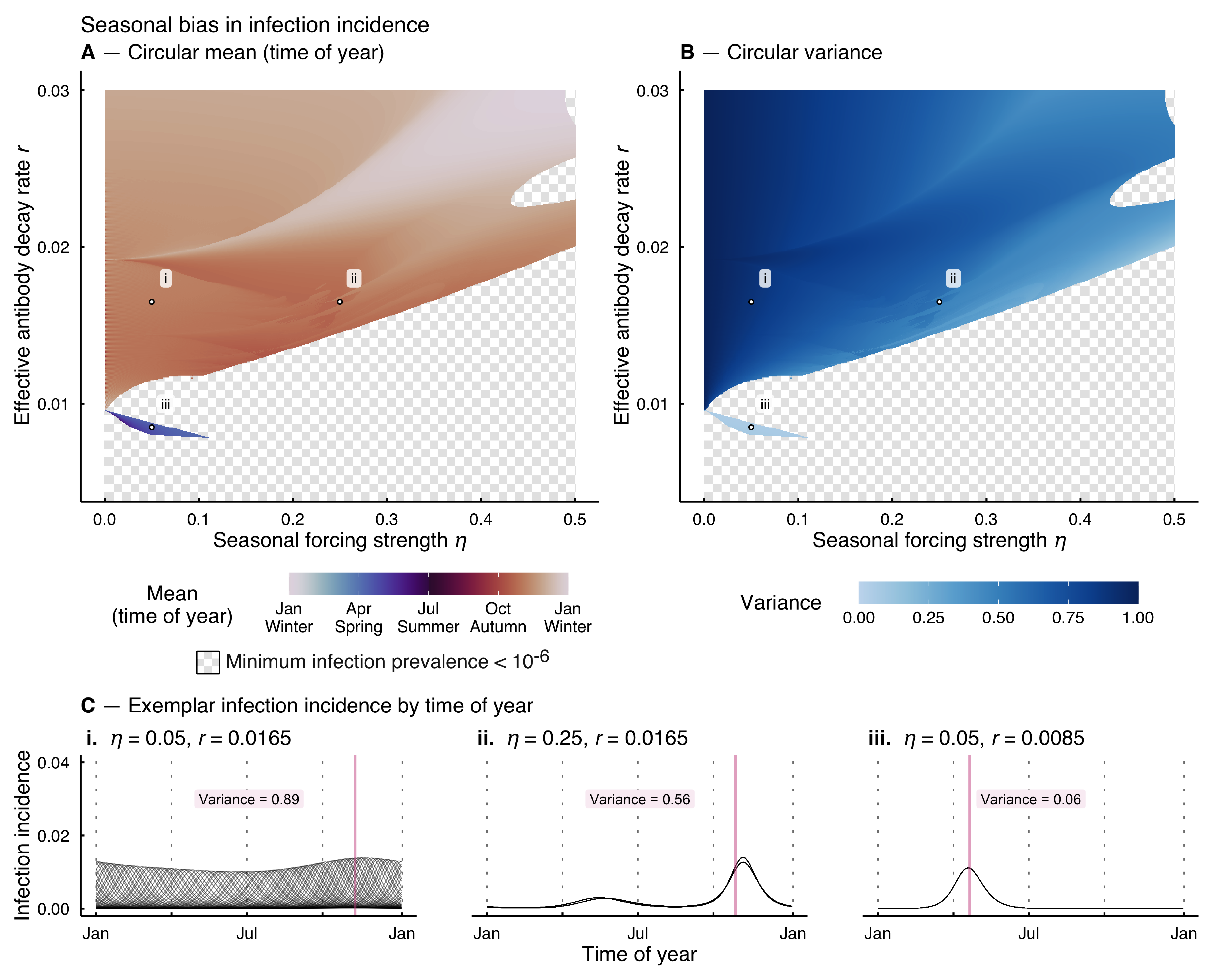}}
     \caption{ The seasonal bias in the time of infection across varied seasonal forcing strength $\eta$ and effective antibody decay rate $r$. Parameters are otherwise as in Table \ref{table-params}. Calculation of the circular mean and variance is detailed in Appendix D. The system was evaluated across a period of 250 years, following 1,000 years of burn-in. Results from parameter combinations that produce a minimum infection prevalence less than $10^{-6}$ are masked (see Figure \ref{figure-grid-seasonality}). The full unmasked results are presented in Appendix Figure \ref{figure-supp-seasonality-bias}. \textbf{A}: The circular mean of time of year of infection incidence. \textbf{B}: The circular variance in time of year of infection incidence, which ranges between zero (entirely concentrated at one day of the year) and one (completely uniform across the year). \textbf{C}: Exemplar infection incidence trajectories by time of year (following burn-in), corresponding to the points \textit{i} through \textit{iv} depicted on the panels above. Trajectories are displayed across 60 years following burn-in. For each model solution, we illustrate both the circular mean (vertical line) and circular variance (text).}
    \label{figure-grid-season-bias}
\end{figure}

The exemplar dynamics described above demonstrate that the timing of infection incidence across each year may vary substantially across different values of the effective antibody decay rate $r$ and seasonal forcing strength $\eta$. We now examine this further, identifying the distribution of infection timing by time of year across the same parameter space. We summarise this distribution using the circular mean and variance, quantities which account for the cyclic nature of the calendar year (Appendix D). The mean timing of infection is most often located around the end of the year (Figure \ref{figure-grid-season-bias}A), while a small region of parameter space yields a later mean infection timing around April (e.g.\ Figure \ref{figure-grid-season-bias}C panel \textbf{iii}).

The circular variance also varies substantially across the parameter space (Figure \ref{figure-grid-season-bias}B). The lowest (unmasked) circular variance is observed for the small region of parameters with a mean infection timing around April. Substantially lower variance can be observed in regions of parameter space that have been masked due to low infection prevalence (Appendix Figure \ref{figure-supp-seasonality-bias}), with this lower prevalence being a direct consequence of the timing of infections being highly concentrated to short periods of the year. Outside of the tongues, the circular variance is highly dependent upon the strength of forcing $\eta$. Where the strength of seasonal forcing is high (e.g.\ $\eta \geq 0.25$), the circular variance is intermediate and the apparent seasonality of infections is less pronounced (e.g.\ Figure \ref{figure-grid-season-bias}C panel \textbf{ii}). For low values of forcing, the circular variance approaches one, implying that (in the long-run) dynamics are highly aseasonal and infections with the virus may occur at any time of year (e.g.\ Figure \ref{figure-grid-season-bias}C panel \textbf{i}). 

These results align with the commonly observed seasonality of respiratory viruses in temperate regions, where different viruses are observed to be most prevalent at different times of the year \parencite{Moriyama2020-ez}. For example, the influenza viruses and respiratory syncytial virus are most prevalent during the wintertime, while rhinovirus is most prevalent during autumn and spring \parencite{Moriyama2020-ez, RedBook-2021}. Although such differences could arise due to differences in the rate at which the different viruses are transmitted across the course of each year (i.e.\ different seasonal forcing functions) or interference between the pathogens \parencite{Nickbakhsh2019-qx}, our results show that large differences in the timing of infections throughout the year could arise solely from differences in the human immune response to the virus or the ability of the virus to evade prior immunity.

\subsection*{Resonant damping of infection incidence}

\begin{figure}[H]
    \centering
    \makebox[\textwidth][c]{\includegraphics[width=13cm]{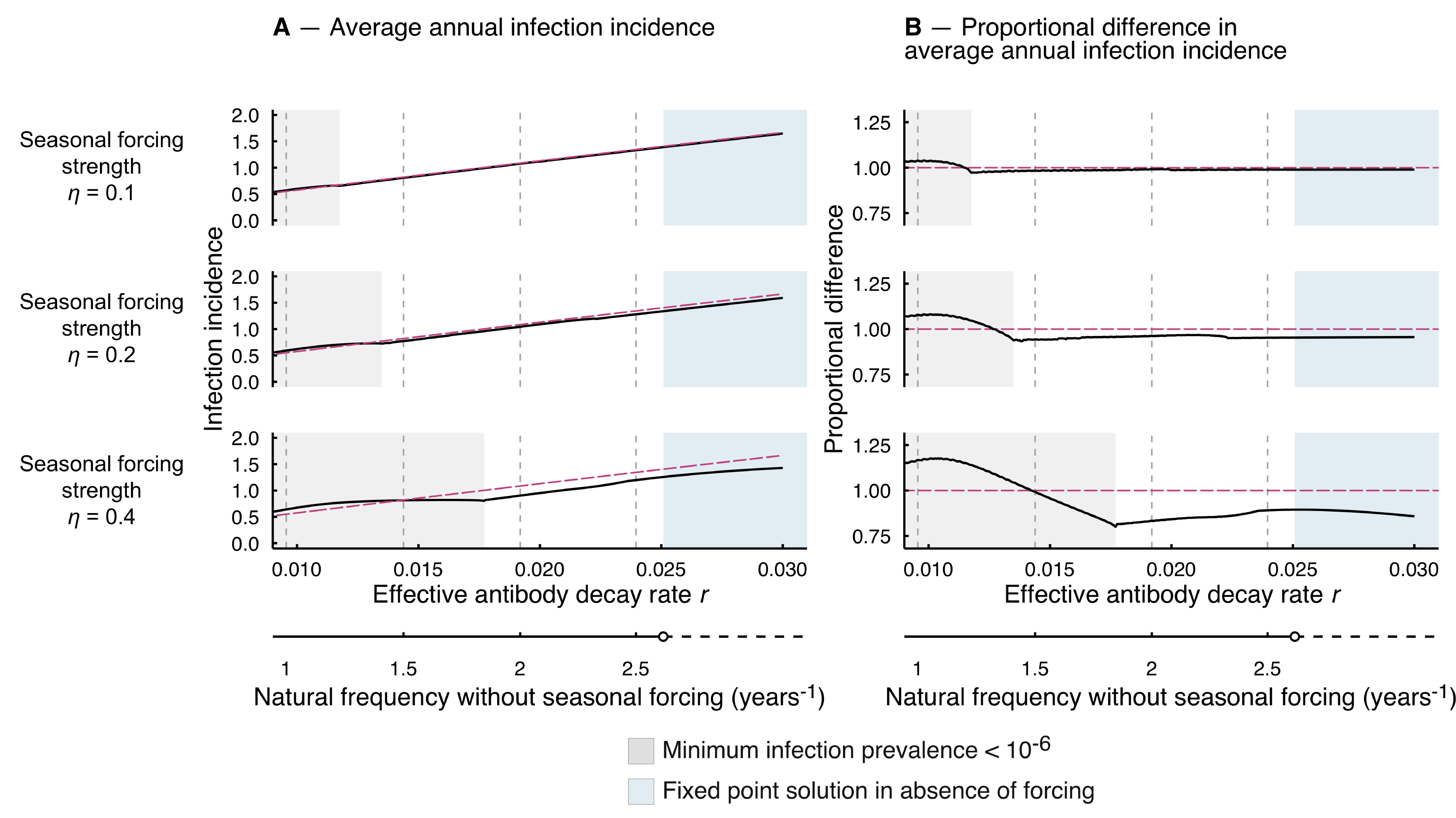}}
     \caption{Changes in average infection incidence due to resonant dynamics induced by seasonal forcing, across varied effective antibody decay rate $r$. All other parameters as listed in Table \ref{table-params}. The corresponding natural frequency of the system (i.e.\ without seasonal forcing) is displayed as a secondary axis. Light blue shading illustrates values of effective antibody decay rate that would go to a fixed point solution in the absence of seasonal forcing, while grey shading indicates values of the effective antibody decay rate that produce a minimum infection prevalence below $10^{-6}$. The system was evaluated across a period of 250 years, following 1,000 years of burn-in. \textbf{A}: The average annual infection incidence following burn-in across effective antibody decay rate and increasing seasonal forcing strength $\eta$. The dashed pink line indicates the average annual infection incidence in the absence of seasonal forcing ($\eta = 0$, as in Figure \ref{figure-bifurcation}C) \textbf{B}: The proportional difference in the average annual infection incidence following burn-in for the given value of seasonal forcing strength $\eta$ compared to the zero seasonal forcing case ($\eta = 0$).}
    \label{figure-seasonality-resonance}
\end{figure}

In addition to altering the timing of infections, we find that the interaction between seasonal forcing and antibody decay can produce a resonant damping that yields a cumulative infection incidence less than would otherwise occur in the absence of forcing. We illustrate this by calculating the average annual infection incidence (i.e.\ the mean daily incidence in the period following burn-in multiplied by 365) for varying effective antibody decay rate $r$ and seasonal forcing strength $\eta$ (Figure \ref{figure-seasonality-resonance}). For weak seasonal forcing ($\eta = 0.1$) the resultant average annual infection incidence differs little from the baseline. However, as the strength of seasonal forcing $\eta$ is increased to $0.2$ or $0.4$, the annual average infection incidence changes substantially (Figure \ref{figure-seasonality-resonance}). For the high seasonal forcing case of $\eta = 0.4$, the average annual incidence is approximately 25\% less than baseline for an effective antibody decay rate of $r = 0.02 \;\text{days}^{-1}$. At this decay rate, the system has a natural frequency (i.e.\ in the absence of seasonal forcing, $\eta = 0$) of approximately two periods per year (Figure \ref{figure-bifurcation}B), such that forcing is often acting in opposition to the natural oscillatory behaviour of the system. We conversely observe increases in average annual incidence around $r = 0.01 \;\text{days}^{-1}$ (with this having a natural frequency of approximately one period per year, i.e.\ the forcing frequency and the natural frequency are very similar), although this also corresponds to a low minimum infection prevalence (less than $10^{-6}$) in the long-term. As such, resonant increases in infection incidence may be associated with stochastic extinction of the virus.

These results show that the interaction between seasonal forcing and waning immunity not only shapes the timing of infections but can produce substantial differences in the long-term average infection incidence of a respiratory virus. These differences are large enough that models would fail to predict the long-term public health impact of a respiratory virus if they did not capture this interaction. While we do not consider vaccination in our model, the presence of such resonant dynamics could have significant implications for the design of vaccination programs, as recurring annual vaccination programs could act as an additional driver of immuno-epidemiological resonance \parencite{Choisy2006-oo}. 

\section{Conclusions}\label{sec-conclusions}

We have examined the influence of antibody-mediated immunity upon the seasonality of respiratory viruses by constructing a novel model of respiratory virus transmission, where an individual's level of immune protection against (re-)infection is dependent upon their effective antibody level, which decays with some combination of antibody waning and antigenic drift. This allows us to explore the deterministic immuno-epidemiological dynamics of respiratory virus transmission capturing mechanisms of antibody-mediated immunity that would otherwise typically be modelled within statistical and agent-based modelling frameworks \parencite{Hogan2023-ps, Muller2023-eu, Le2024-ku, Conway2024-wq, Yuan2025-le, Hao2025-io}. Using this model, we have found that periodic epidemic waves can arise in the absence of seasonal forcing and have identified two associated Hopf bifurcations across the effective antibody decay rate. This includes a subcritical Hopf bifurcation yielding multi-stability across low effective antibody decay rates. Where seasonal forcing is included in the model, we have performed a comprehensive parameter sweep examining the interaction between the effective antibody decay rate and the strength of seasonal forcing and have shown that diverse dynamical outcomes can arise from this interaction. These include periodic dynamics, which resemble the annual (or multi-annual) seasonal epidemics of endemic respiratory viruses in temperate regions \parencite{Moriyama2020-ez}, and chaotic dynamics, which we have found can still exhibit (in the qualitative sense) seasonal behaviour. We have also identified a regime of quasiperiodicity, with epidemic waves recurring at a regular interval but never aligning with the calendar year in the long-term, a pattern not present in public health surveillance data of respiratory viruses. Using circular statistics to characterise the seasonal dynamics, we have shown that these different dynamical regimes produce varying patterns of infection timing by time-of-year, and this timing varies with the interaction between the effective antibody decay rate and seasonal forcing. Beyond shaping the seasonality of infection timing, we find this interaction can produce a resonant amplification or damping, leading to non-linear changes in the long-term infection incidence as a function of the effective antibody decay rate.

Our modelling study is subject to several limitations that should be considered when interpreting these findings. First, we explore only the deterministic dynamics of the system. Consequently, we have masked results in regions of parameter space where it is highly likely that stochastic extinction could occur (while we chose a threshold of a minimum infection prevalence of less than $10^{-6}$ as in depicted in Figure \ref{figure-grid-seasonality}, other choices would be appropriate depending on the population size considered). Consideration of our compartmental model within a continuous-time Markov chain framework would allow for this extinction risk to be directly investigated and enable us to explore the long-term stability of dynamical phenomena identified in our study such as quasiperiodicity. Second, we do not consider the role of demographic change in our model. While we argue in Section \ref{sec-model} that a simple demographic turnover process (i.e.\ constant birth and death rate across the population) would be conceptually similar to an increase in the effective antibody decay rate, an age-stratified model would (of course) produce distinct quantitative results and perhaps distinct qualitative results given the substantial heterogeneity in the age-specific transmission patterns and disease burden for respiratory viruses \parencite{Hogan2016-ci, Lau2020-il, Zimmerman2022-kj, Eales2023-ww, Boldea2024-pt}. Finally, our use of a constant effective antibody decay does not capture the possibility of punctuated evolution \parencite{Roberts2019-uh} or the rate of evolution being dependent upon population immunity \parencite{Boni2006-rz, Lange2009-pc, Bedford2015-zb}. Such effects could be captured within our antibody-structured framework to explore how they affect the immuno-epidemiological dynamics.

The modelling framework we have presented could be extended to capture the concurrent transmission of multiple viral strains where each strain imparts a cross-reactive antibody response. This kind of cross-reactive immunity, typically arising as a consequence of antigenic shift (rather than the antigenic drift we have captured with the effective antibody decay rate here), can be found among many respiratory viruses \parencite{Sullender2000-kj, Carter2013-bx, Koutsakos2023-hi}. Such an extension would involve the inclusion of multiple infected classes and a (multi-dimensional) stratification of the susceptible class by the antibody level imparted by each strain, with this allowing for immune protection due to prior infection to be specified in terms of antibody cross-reactivity (unlike previously published multi-strain models that use relative cross-protection values, e.g.\ \cite{Gupta1996-ct, Andreasen1997-ou, Kucharski2016-kf}). This multi-strain model could be used to explore, for example, the transmission dynamics of influenza, for which cross-reactivity between sub-types of influenza is likely a key driver \parencite{Andreasen1997-ou, Carter2013-bx, Krammer2019-nd}. Such a multi-strain model could also be used to explore the phenomenon of antibody-dependent enhancement, where the presence of cross-reactive antibodies can lead to a heightened susceptibility to infection or worsened disease outcomes \parencite{Wells2025-jb}. A stochastic implementation of this framework would allow for exploration of how cross-reactive immunity affects the likelihood of emergence and sustained transmission of a novel strain of a virus.

Investigating the dynamics of any particular respiratory virus with our model would require careful parametrisation of transmission, immunity and seasonal forcing. Given appropriate parametrisation, such an analysis would enable the exploration of the biological plausibility of the different dynamic behaviours that we have highlighted (such as quasiperiodicity or resonant amplification). Further, such a parametrised study could be of particular value in application to SARS-CoV-2. While early modelling studies predicted that the virus would, in the long-term, approach a stable pattern of once-a-year winter seasonality in temperate regions (e.g.\ \cite{Kissler2020-aq, Townsend2023-xg}), the virus does not yet appear to have fallen into this pattern in many temperate regions, with off-winter waves continuing to be observed \parencite{Donovan2025-zu, Rubin2025-du, UK-Health-Security-Agency2025-wj}. Our results highlight how a variety of seasonal patterns beyond that of once-a-year winter seasonality --- such as quasiperiodic waves that do not align with the calendar year --- could emerge as the long-term behaviour of a respiratory virus like SARS-CoV-2 as a natural consequence of the interaction between the periodic drivers of waning immunity and seasonal forcing.

\section{Acknowledgements}

We thank Cameron Zachreson for their helpful input on this research. Ruarai Tobin was supported by a Melbourne Research Scholarship. This research was also supported by The University of Melbourne’s Research Computing Services and the Petascale Campus Initiative. 

\section{Funding}

James McCaw was supported by an ARC Laureate Fellowship (FL240100126). Freya Shearer was supported by an NHMRC Investigator Grant (Emerging Leader Fellowship, 2021/GNT2010051). These funders had no role in study design, collection, analysis and interpretation of data, writing of the report or decision to submit the article for publication.

\printbibliography

\pagebreak

\section*{Appendix}

\textbf{Appendix Figures}

\setcounter{figure}{0}
\renewcommand{\figurename}{Appendix Figure}
\begin{refsection}

\begin{figure}[H]
    \centering
    \makebox[\textwidth][c]{\includegraphics[width=10cm]{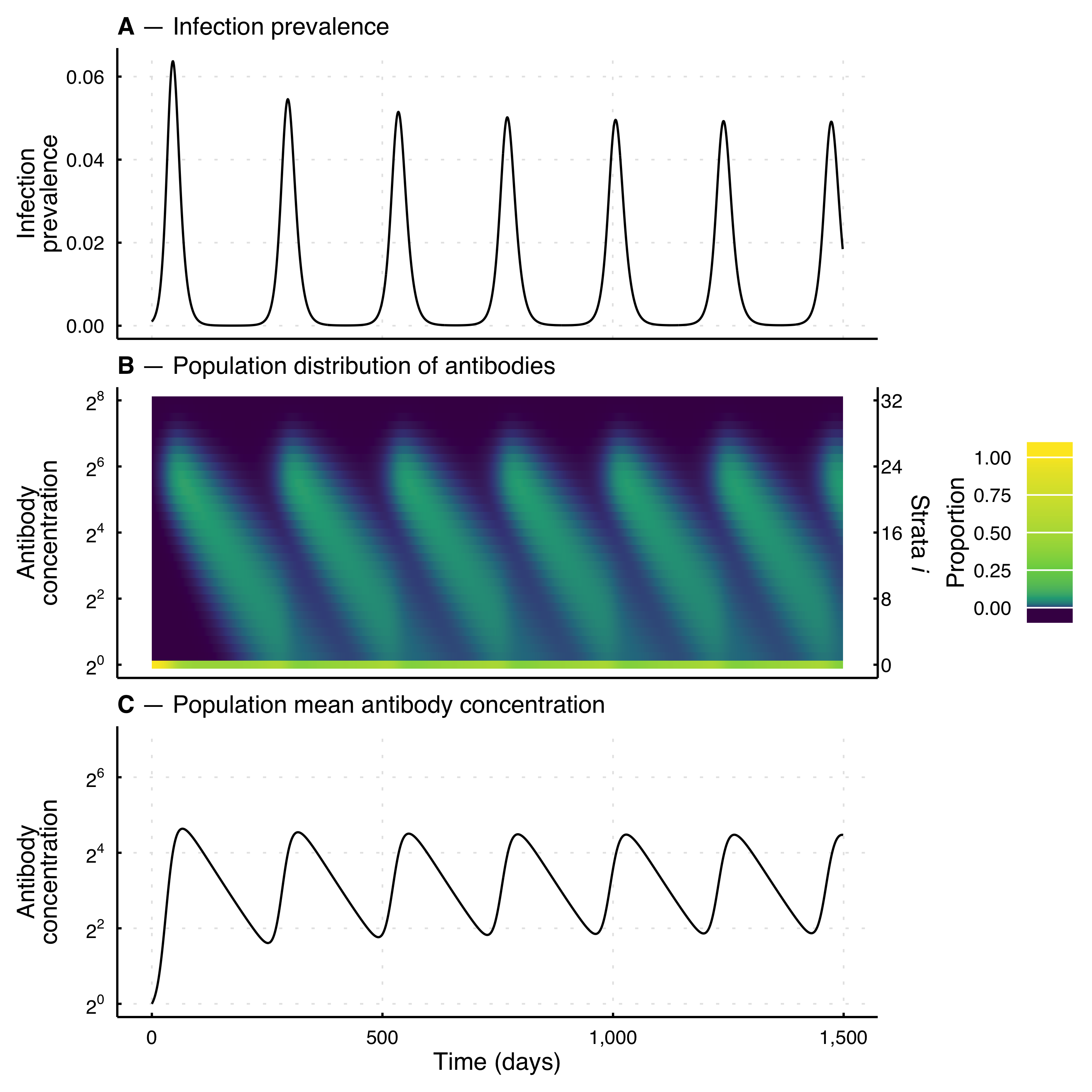}}
     \caption{Characteristic dynamics of the immunity-structured model of respiratory virus transmission. Model parameters are as in Table 1. \textbf{A}: Prevalence (i.e.\ proportion) of infectious individuals $I$ over time. \textbf{B}: Proportion of susceptible individuals in each strata $S_i$ for $0 \leq i \leq k$. \textbf{C}: The mean antibody concentration across the susceptible population, calculated as an arithmetic mean across concentrations (i.e.\ not a geometric mean).}
    \label{figure-supp-basic}
\end{figure}

\begin{figure}[H]
    \centering
    \makebox[\textwidth][c]{\includegraphics[width=8cm]{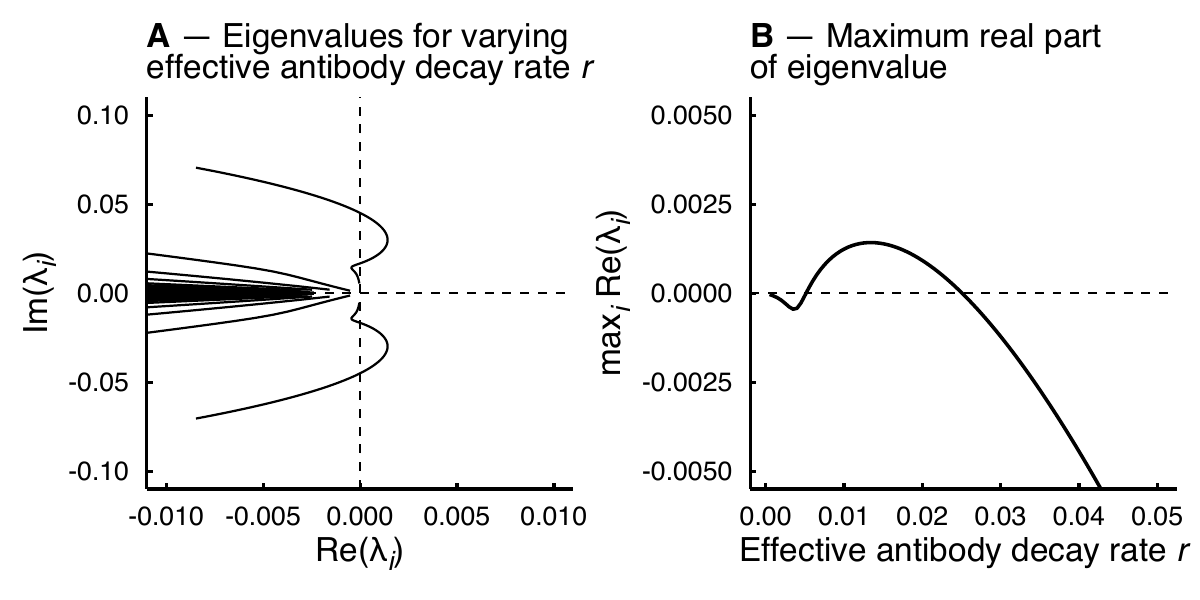}}
     \caption{Eigenvalues of the linearised ODE system at the fixed point solution as antibody decay rate $r$ is varied from $0$ to $0.15$. The linearised system was reduced by taking $I = 1-\sum_{i=0}^kS_i$. \textbf{A}: Path of the eigenvalues $\lambda_i$ as decay rate is varied. \textbf{B}: Maximal real part of the eigenvalues. Two Hopf bifurcations occur as this value crosses the x-axis. }
    \label{figure-supp-eigenvalues}
\end{figure}

\begin{figure}[H]
    \centering
    \makebox[\textwidth][c]{\includegraphics[width=7cm]{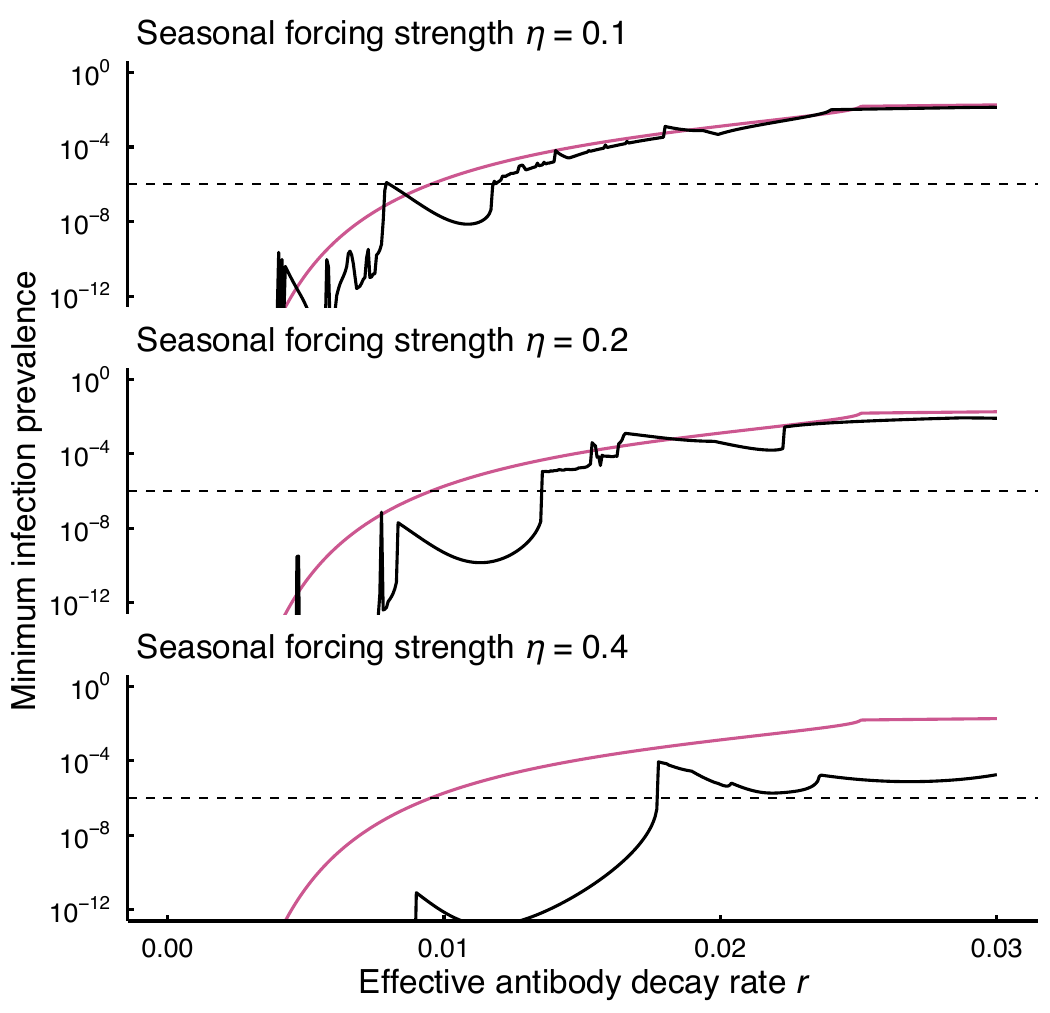}}
     \caption{The minimum infection prevalence observed for varying antibody decay rate $r$ and strength of seasonal forcing $\eta$, where all other parameters are as in Table 1. The black line indicates the minimum infection prevalence for the specified value of $\eta$, while the red line indicates the minimum infection prevalence in the absence of seasonal forcing ($\eta = 0$). The horizontal dashed line indicates the threshold of $10^{-6}$ used elsewhere as a threshold of potential stochastic extinction. The system was evaluated across a period of 250 years, following 1,000 years of burn-in.}
    \label{figure-supp-seasonality-extinction}
\end{figure}

\begin{figure}[H]
    \centering
    \makebox[\textwidth][c]{\includegraphics[width=13cm]{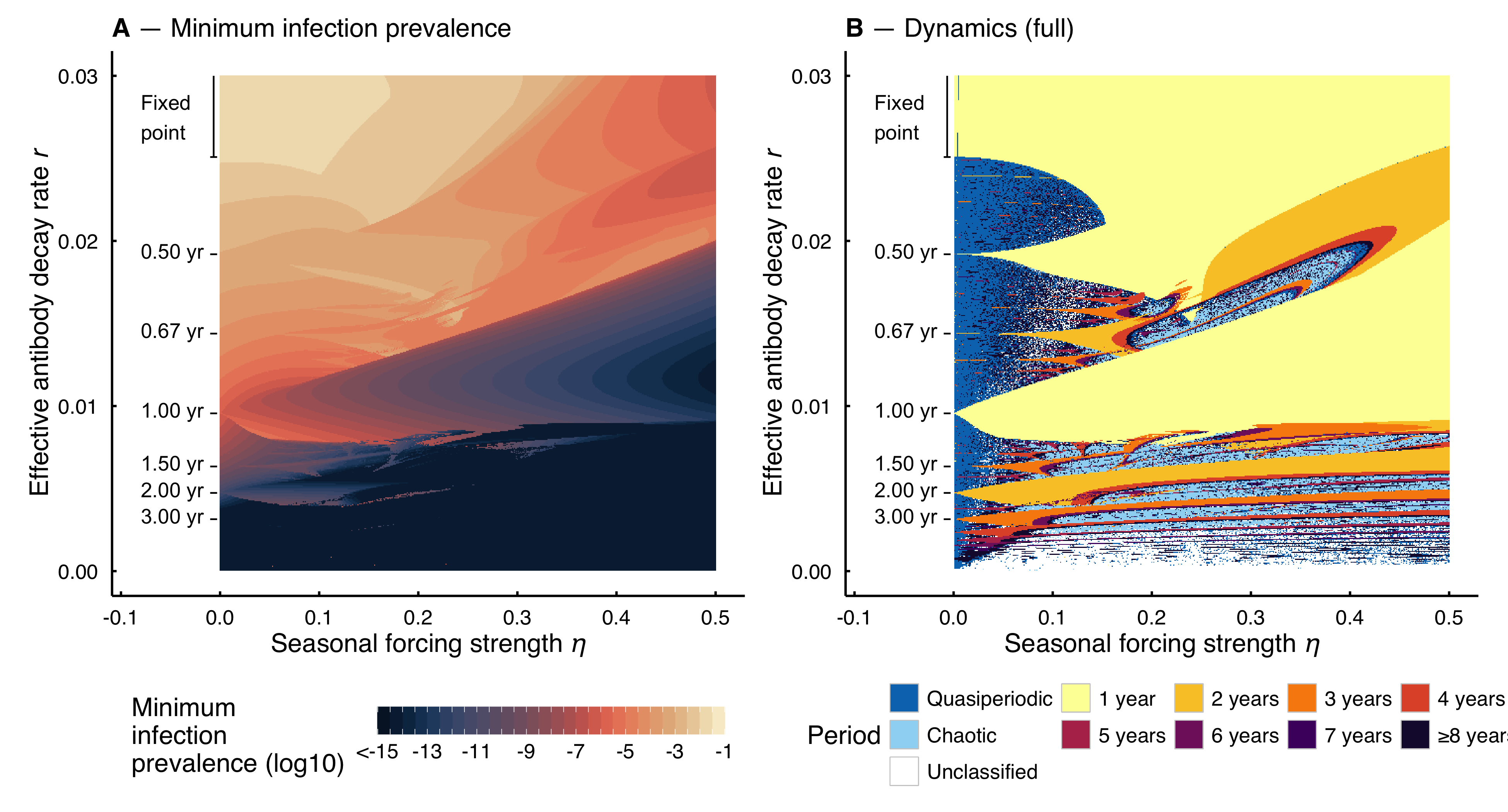}}
     \caption{The results presented in Figure \ref{figure-grid-seasonality} with low minimum infection prevalence results not masked. For full details, see Figure \ref{figure-grid-seasonality}. }
    \label{figure-supp-seasonality-full}
\end{figure}

\begin{figure}[H]
    \centering
    \makebox[\textwidth][c]{\includegraphics[width=14cm]{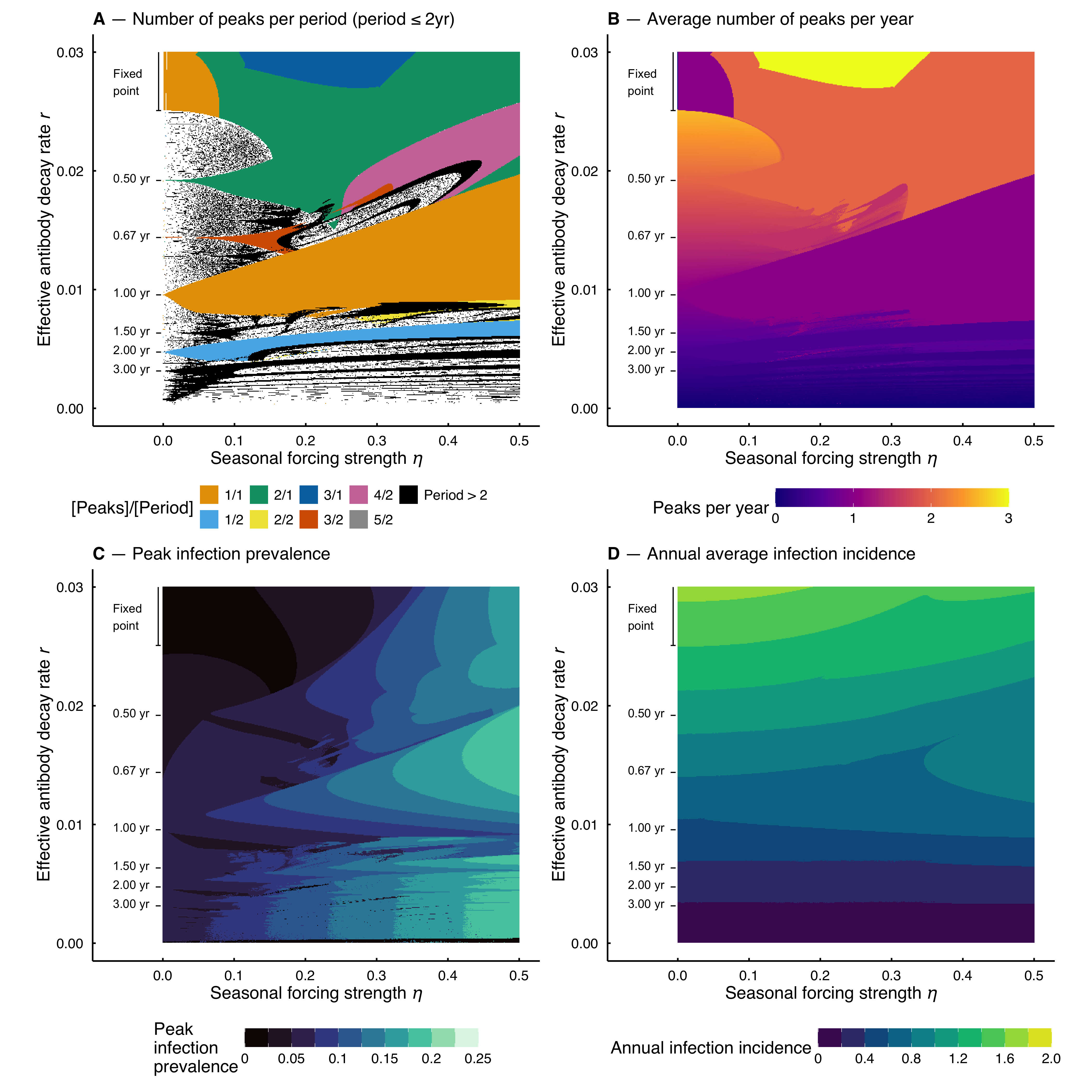}}
     \caption{Differences in model dynamics across varied seasonal forcing strength $\eta$ and antibody decay rate $r$. Model parameters are otherwise as in Table 1. Extends Figure 4 in the main text. The system was evaluated across a period of 250 years, following 1,000 years of burn-in. As in Figure 4, the natural period of the system in the absence of seasonal forcing ($\eta = 0$) is indicated on the left-hand side of each plot. \textbf{A}: The number of peaks per period duration (where period is at most two years). \textbf{B}: The average number of peaks per year across the 250 years following burn-in. \textbf{C}: Peak infection prevalence across the 250 years following burn-in. \textbf{D}: The average annual infection incidence across the 250 years following burn-in. }
    \label{figure-supp-seasonality}
\end{figure}

\begin{figure}[H]
    \centering
    \makebox[\textwidth][c]{\includegraphics[width=13cm]{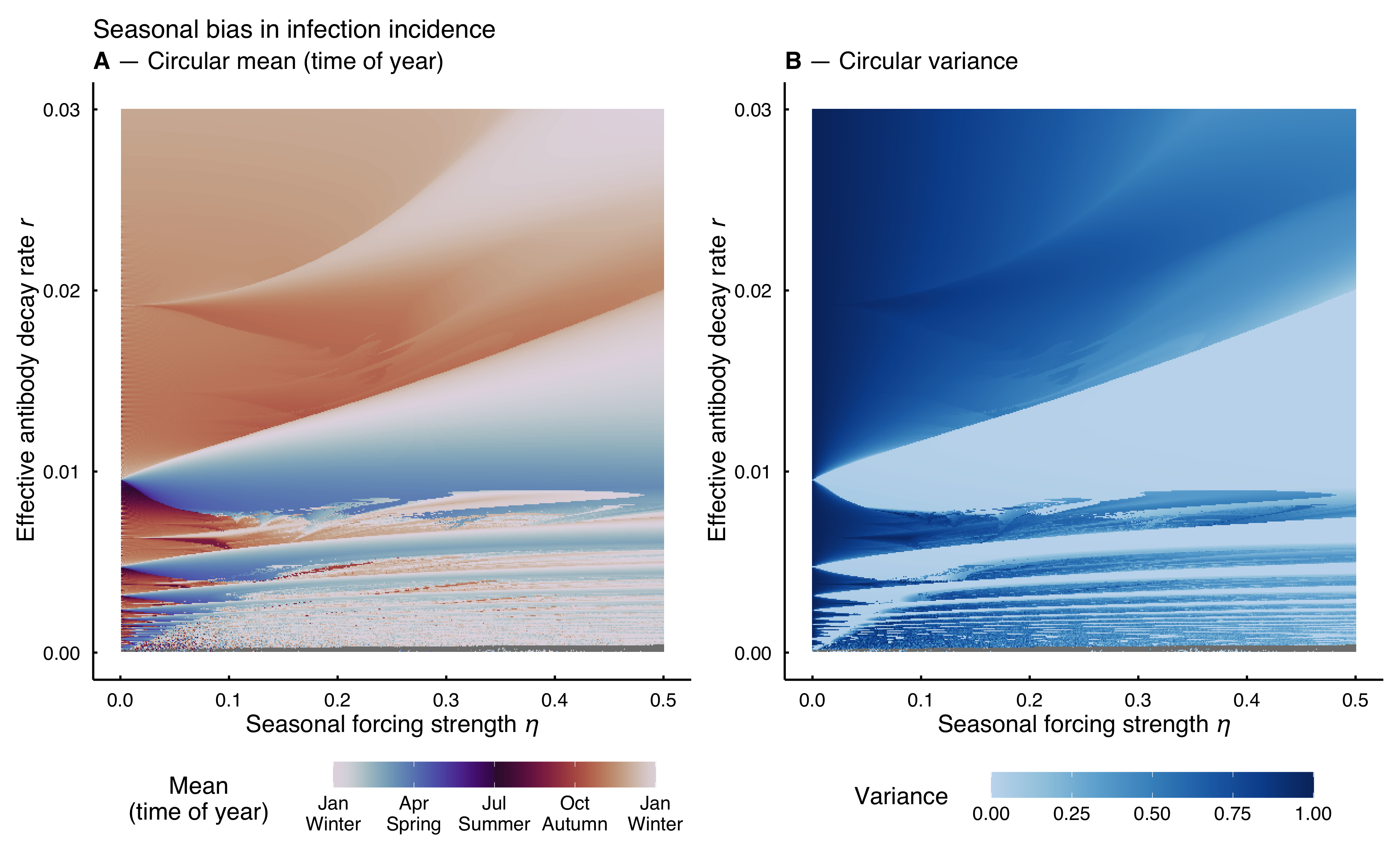}}
     \caption{ The results presented in Figure \ref{figure-grid-season-bias} with low minimum infection prevalence results not masked. For full details, see Figure \ref{figure-grid-season-bias}. }
    \label{figure-supp-seasonality-bias}
\end{figure}

\begin{figure}[H]
    \centering
    \makebox[\textwidth][c]{\includegraphics[width=7cm]{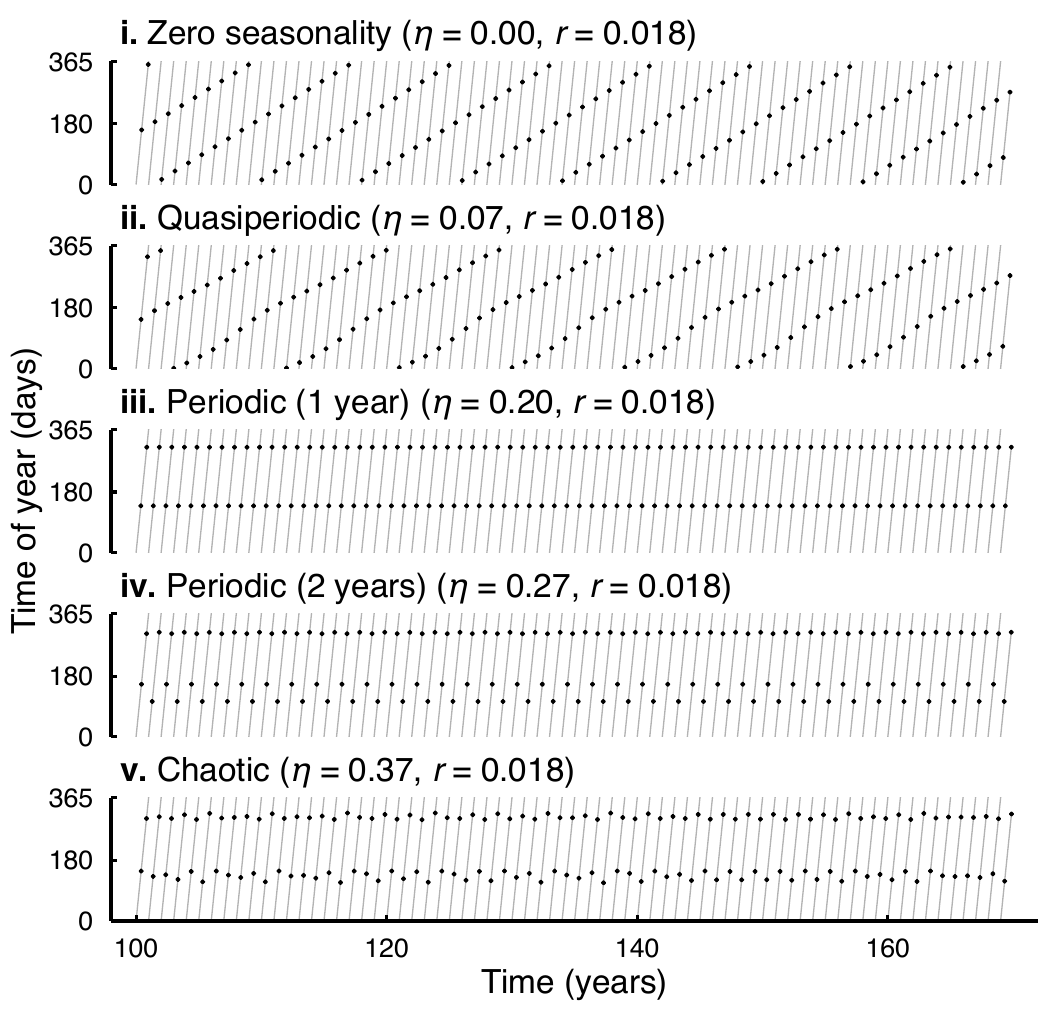}}
     \caption{Timing of peak infection incidence (following 100 years of burn-in) for five exemplar sets of parameters (the same as those in Figure 5). Points indicate the timing of peak infection incidence by time of year (y-axis) and year (x-axis). Diagonal lines indicate the progression of time. }
    \label{figure-supp-peak-timing}
\end{figure}

\clearpage
\section*{Appendix A}

\textbf{Compartmental discretisation for the exponential decay in antibody concentration}

Here, we derive the compartmental model of exponential decay in the susceptible population which is used in Section 2. We assume that individuals have some antibody concentration $c(t)$ which decays exponentially over time with decay rate $r$:
\begin{equation*}
    \dv{c}{t} = -r c(t).
\end{equation*}
This may be equivalently stated as a linear decay in the log-2-transformed concentration:
\begin{equation}\label{eq-linear-decay}
    \dv{\log_{2}(c)}{t} = -\frac{r}{\log_e(2)}.
\end{equation}
For the sake of argument, this this process may be described across the population of susceptible individuals using a partial differential equation. Let $S(t, \log_{2}(c))$ be the fraction of susceptible individuals at time $t$ who have a log-2 antibody concentration of $\log_{2}(c)$. From Equation \ref{eq-linear-decay}, we have:
\begin{equation}\label{eq-pde}
    \pdv{S}{t} - \frac{r}{\log_e(2)}\pdv{S}{\log_{2}(c)} = 0
\end{equation}
This PDE can then be approximated using the method of lines (e.g.\ \cite{Plant1986-jo}). First, we discretise antibody concentration across discrete strata $i$, taking:
\begin{equation}\label{eq-concentration-supp}
    c_i = 2^{a(i/k)},
\end{equation}
where $0 \leq i \leq k$ (such that $2^0 \leq c_i \leq 2^a$). The partial derivative of $S$ with respect to $\log_{2}(c)$ is then approximately:
\begin{equation*}
    \pdv{S}{\log_{2}(c)} \approx \frac{S_{i+1} - S_i}{\log_{2}(c_{i+1}) - \log_{2}(c_i)} = \frac{S_{i + 1} - S_i}{a/k}
\end{equation*}
Substituting this approximation into our partial differential equation (Equation \ref{eq-pde}) yields:
\begin{equation*}
    \dv{S_i}{t} = \frac{r k}{a \log_e(2)}(S_{i+1}-S_i)
\end{equation*}
To simplify the equation, we take:
\begin{equation*}
    \rho = \frac{r}{a\log_{e}2},
\end{equation*}
such that we have
\begin{equation*}
    \dv{S_i}{t} = \rho k(S_{i+1}-S_i).
\end{equation*}
This yields our compartmental discretisation of the antibody decay process, with individuals in compartment $S_{i+1}$ transitioning to compartment $S_i$ at rate $\rho k$:
\begin{equation*}
\boxed{S_{i+1}} \xrightarrow[\mbox{\normalsize$\rho k$}]{} \boxed{S_{i}}
\end{equation*}
for $i > 0$. 

The parameter $k$ controls the resolution of our discretisation of Equation \ref{eq-pde}, with increasing $k$ reducing the variance (dispersion) in the antibody decay process across the population. To show this, we note that the probability of an individual having made $i$ transitions by time $t$ is given by the Poisson distribution $I(t) \sim \text{Poisson}(\rho k t)$ (where we are ignoring the possibility of reaching the minimum strata). Assuming that all individuals in the population start at $t = 0$ with the same antibody concentration, the change in $\log_{2}(c(t))$ across the population following $I(t)$ transitions is then (using Equation \ref{eq-concentration-supp}):
\begin{equation*}
    -\frac{aI(t)}{k},
\end{equation*}
which is distributed with a variance of:
\begin{equation*}
    \text{Var}\left(-\frac{aI(t)}{k}\right) = \frac{a^2\rho}{k}t.
\end{equation*}
As such, increasing $k$ reduces the variance introduced by discretisation.

\section*{Appendix B}

\textbf{Transmission model with stratification across the infectious class.}

Here, we consider a formulation of our transmission model where we stratify both the susceptible and infectious class by their serum antibody concentration (i.e.\ not only the susceptible class). This allows us to specify a post-infection antibody concentration with dependence upon the pre-infection antibody concentration. We use this to examine antibody boosting, where recovery from infection leads to a linear increase in the log-transformed antibody concentration. 

This model has $2(k + 1)$ compartments:
\begin{equation*}
    \left\{S_0,\;S_1,\;\ldots,\;S_k,\;I_0,\;I_1,\;\ldots,I_k\right\},
\end{equation*}
and the process of antibody waning occurs as in the original model, with:
\begin{equation*}
\boxed{S_i} \xrightarrow[\mbox{\normalsize$(1 - \omega_i)\beta I$}]{} \boxed{I} 
\end{equation*}

Infection occurs in a similar manner to the original model, although the force of infection is now calculated across the sum of the infectious compartments, i.e. $\beta \sum_{i=0}^kI_i$.  Upon infection, an individual in $S_i$ transitions to compartment $I_i$ (i.e.\ the same strata index $i$):
\begin{equation*}
\boxed{S_i} \xrightarrow[\mbox{\normalsize$(1 - \omega_i)\beta \sum_{i=0}^kI_i$}]{} \boxed{I_i} 
\end{equation*}
where $\omega_i$ is the level of protection against infection (given exposure) for strata $i$, as defined in the main text.

Upon recovery from infection, an individual in compartment $I_i$ may transition to any compartment $S_j$ where $j \geq i$. This occurs according to the probability $P(I_i \rightarrow S_j)$, i.e.:
\begin{equation*}
\boxed{I_i} \xrightarrow[\mbox{\normalsize$\gamma P(I_i \rightarrow S_j)$}]{} \boxed{S_j} 
\end{equation*}
As in the original model, this probability may be defined to capture different mechanisms. Here, we consider a model of antibody boosting, where the log-concentration of antibodies following infection $\log_{2}(c_\text{post})$ increases linearly with the pre-infection concentration $\log_{2}(c_\text{pre})$, i.e.:
\begin{equation*}
    \log_{2}(c_\text{post}) = \log_{2}(c_\text{pre}) + X
\end{equation*}
where $X \sim N(\mu, \sigma)$ is a Normal distribution defining the jump in log-concentration. This defines the probabilities $P(I_i \rightarrow S_j)$:
\begin{equation*}
    P(I_i \rightarrow S_j) = \begin{cases}
        P(a \leq X), \quad &j = k,\\
        P\left(a\frac{i}{k} < X \leq a\frac{i+1}{k}\right), \quad &i > i \text{ and } j < k,\\
        P\left(X < a\frac{i+1}{k}\right), \quad &j = i,\\
    \end{cases}
\end{equation*}
where we have applied the additional constraints that antibody concentration cannot decrease and may not be above $2^a$. In the below results, we have taken $X \sim N(6, 0.5)$, such that the distribution of post-infection antibody concentration is approximately the same as in the main text where pre-infection antibody concentration is minimal ($2^0$).

In total, we have the following dynamical system defined by $2(k+1)$ ordinary differential equations:
\begin{align*}
    \dv{S_i}{t} &= \rho k ([i < k]S_{i + 1} - [i > 0]S_i) - \beta (1-\omega_i) S_i \sum_{j=0}^k I_j + \gamma\sum_{j=0}^k I_j P(I_j \rightarrow S_i)\\[1em]
    \dv{I_i}{t} &= \beta (1 - \omega_i) S_i \sum_{j=0}^k I_j - \gamma I_i,\\
\end{align*}
which is illustrated in Appendix Figure \ref{compartmental-model-boosting}.

\begin{figure}[H]
    \centering
    \makebox[\textwidth][c]{\includegraphics[width=12cm]{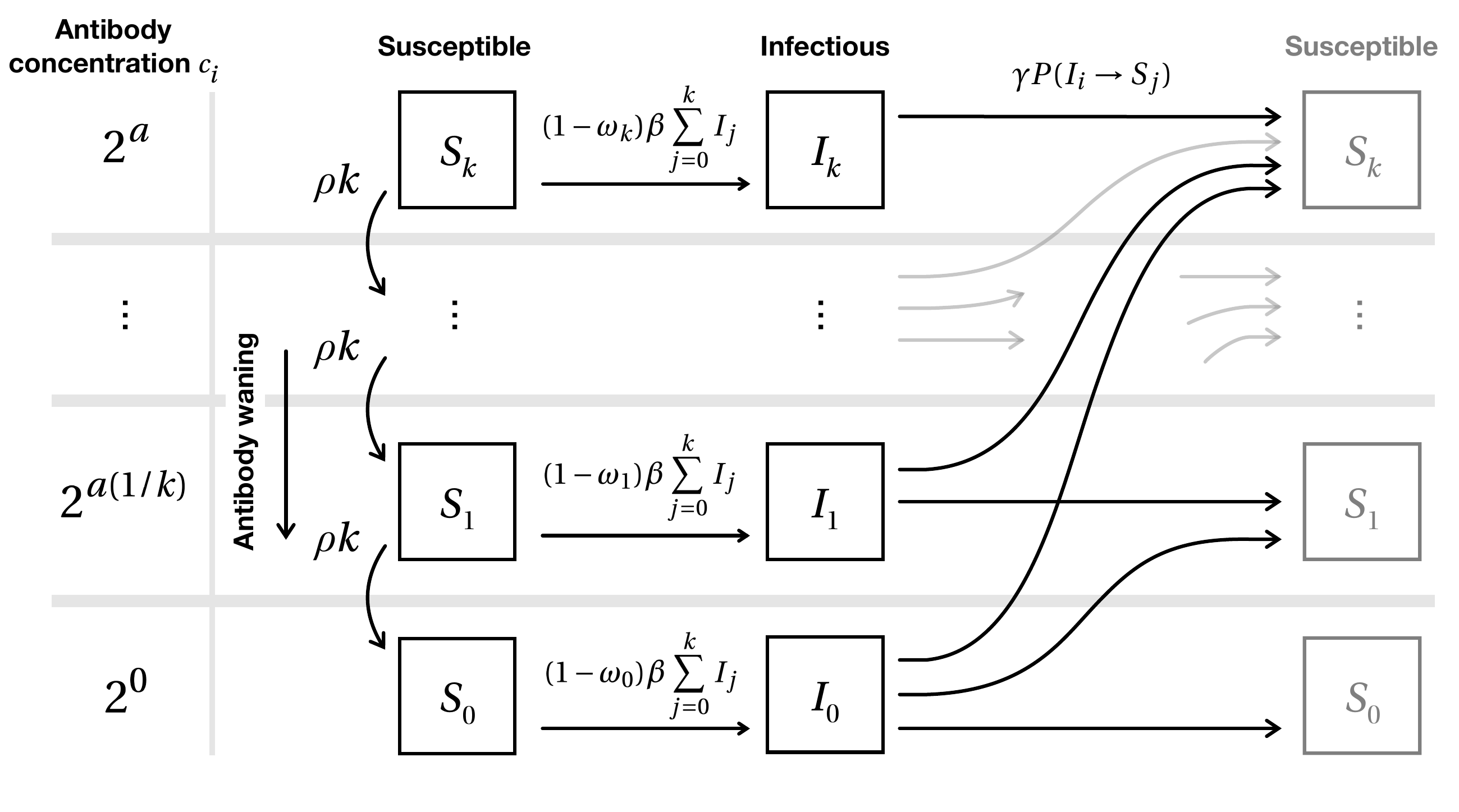}}
     \caption{Diagram of the immuno-epidemiological model with stratification across both the susceptible and infectious classes.}
    \label{compartmental-model-boosting}
\end{figure}

Here, we compare the behaviour of this model with antibody boosting to that of the original model in the absence of seasonal forcing (Appendix Figure \ref{figure-boosting-results}). The bifurcation diagram presented in Appendix Figure \ref{figure-boosting-results}A illustrates a slight difference in the location of the fixed point and periodic solutions (and accompanying bifurcations). Similarly, both the frequency of the periodic solutions and the average annual infection incidence following burn-in are slightly lower for the model with boosting than the original model (Appendix \ref{figure-boosting-results}B). This similarity is also apparent in the exemplar trajectories of infection prevalence (Appendix Figures \ref{figure-boosting-results}D), while clearer differences can be seen in the underlying population average antibody concentration (Appendix Figure \ref{figure-boosting-results}E).

\begin{figure}[H]
    \centering
    \makebox[\textwidth][c]{\includegraphics[width=14cm]{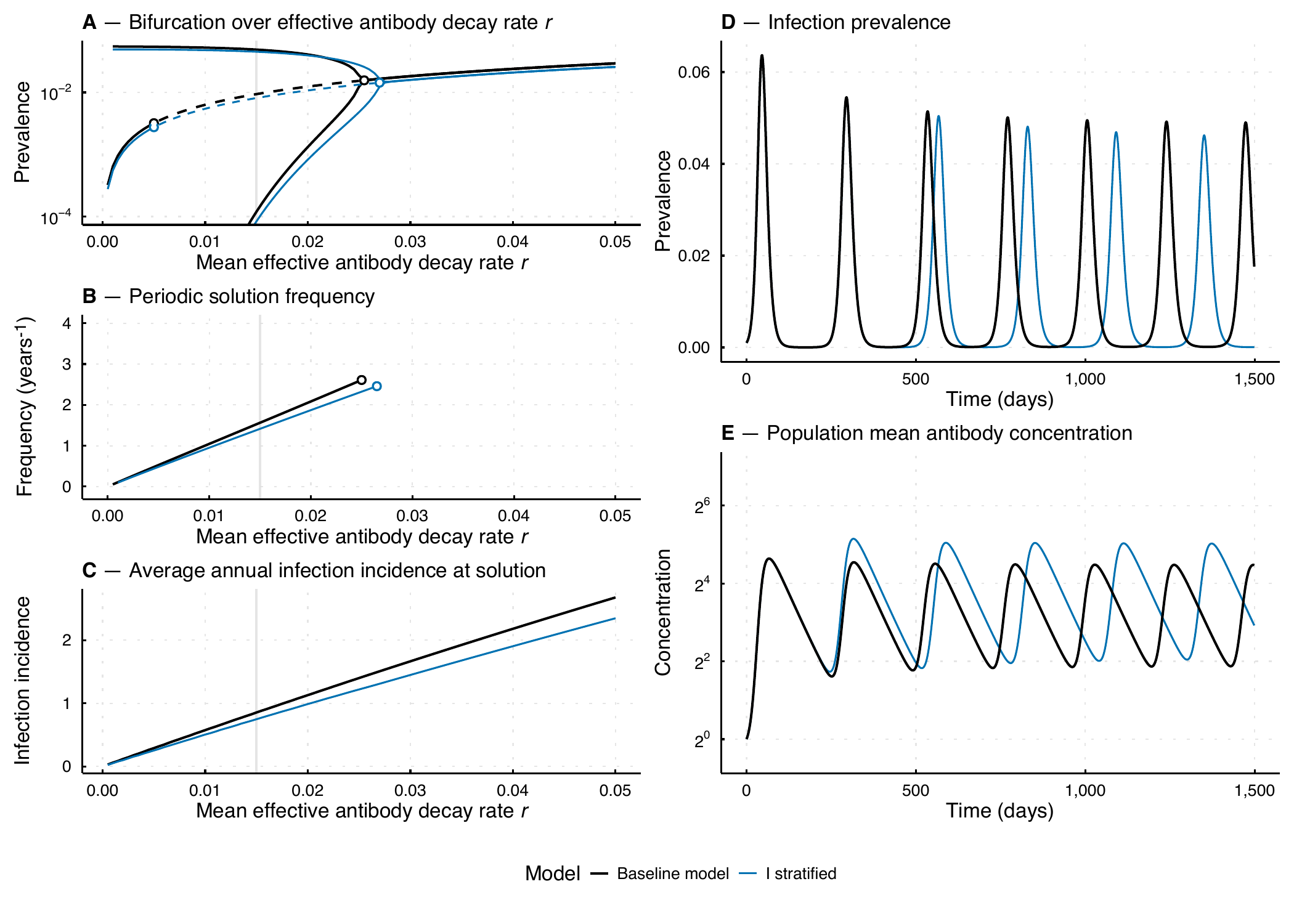}}
     \caption{Dynamics of the immuno-epidemiological model for varying values of antibody decay rate $r$ (all other model parameters are as in Table 1 The system was evaluated across a period of 100 years (36,500 days), following 100 years of burn-in. \textbf{A}: Bifurcation diagram over varying antibody decay rate $r$, with lines depicting infection prevalence at the model solutions (noting the log scale on the y-axis). For the periodic solutions, the light blue lines correspond to the maximal or minimal infection prevalence across each solution. For low values of antibody decay rate, the minimal infection prevalence rapidly approaches values where stochastic effects would be dominant in realistic settings and our deterministic results must be interpreted with care. The separatrix between the stable limit cycle and the stable fixed point for values of $r$ less than approximately $0.005$ is not depicted as it could not be identified numerically. \textbf{B}: The frequency (in years$^{-1}$) of the periodic model solutions for varying antibody decay rate $r$. \textbf{C}: The average annual infection incidence at the model solution (fixed point or stable periodic) for antibody decay rate $r$, calculated as the mean infection incidence multiplied by 365. These average annual infection incidences correspond to both the periodic and (unstable or stable) fixed point solutions. \textbf{D}: Exemplar infection prevalence $I(t)$ for the two models, where we have taken $r = 0.015$. \textbf{E}: Exemplar population mean antibody concentration for the same system as in panel D. }
    \label{figure-boosting-results}
\end{figure}

\section*{Appendix C}

\textbf{Algorithm to determine qualitative dynamics}

In this section, we describe the numerical method we used to determine the qualitative dynamics of a solution, producing (approximate) classifications across between periodicity, quasiperiodicity and chaos. We also provide our method for determining the period length of a periodic solution. 

We first attempt to determine the period length (where it exists). Let $\mathbf{x}(t) = \{S_0(t), S_1(t), \ldots, S_k(t), I(t)\}$ be the state of our system at time $t$. We consider the system for time-points $t$ separated by a time-step of $\Delta t$ between $t_0$ and $t_\text{max}$, i.e.\
\begin{equation*}
    t \in [t_0, t_0 +\Delta t, t_0 + 2\Delta t, \ldots, t_\text{max}]
\end{equation*}
where $t_0$ is a time-point such that model transients are minimal (i.e.\ following some burn-in period) and $t_\text{max}$ is the final time-point of our model solution. For our results, we have taken a time-step of $\Delta t = 0.25$.

By definition, in the absence of external forcing (i.e.\ seasonal forcing), at least one period will have occurred between the times $t_0$ and $t_i > t_0$ if $\mathbf{x}(t_0)=\mathbf{x}(t_i)$. To account for numerical error in our solutions of $\mathbf{x}(t)$, we take this as occurring where:
\begin{equation*}
    ||\mathbf{x}(t_i) - \mathbf{x}(t_0)|| < \epsilon
\end{equation*}
where $\epsilon = 10^{-6}$ for this manuscript. This condition is illustrated for an example model solution in Appendix Figure \ref{figure-ex-period-alg}.

To identify period length, we first identify the time-points $t_0 < t_i <t_\text{max}$ for which the above condition is fulfilled. Where we identify consecutive time-points, we replace each group of consecutive time-points with their central (mean) value. Let $T = [t_1, \ldots, t_n]$ be the ordered array of time-points we identify in this manner. This is then used to calculate samples of period duration $p_i$:
\begin{equation*}
    p_i = \frac{t_i - t_0}{i},
\end{equation*}
with the mean of these samples forming our estimate of the period $\bar{p}$.

In the absence of seasonal forcing, we define a periodic solution as occurring where we have at least two samples of $p_i$ and the standard deviation across $p_i$ is less than one. 

In the presence of seasonal forcing, we define a periodic solution as occurring where we have at least two samples of $p_i$ and the mean period $\bar{p}$ is such that the corresponding frequency is approximately a harmonic of 365 days, i.e.:
\begin{equation*}
    \exists n \in \{1, 2, \ldots, 20\} \quad \text{such that}\quad \min \left\{\text{mod}(n\cdot\bar{p}, 365),\;\; 365- \text{mod}(n \cdot \bar{p}, 365)\right\} < 1.
\end{equation*}
The factor $n$ allows us to periods at a minimum of $1/20$ years. However, where seasonal forcing was present, we did not identify any such sub-annual periodic solutions.

\begin{figure}[H]
    \centering
    \makebox[\textwidth][c]{\includegraphics[width=7cm]{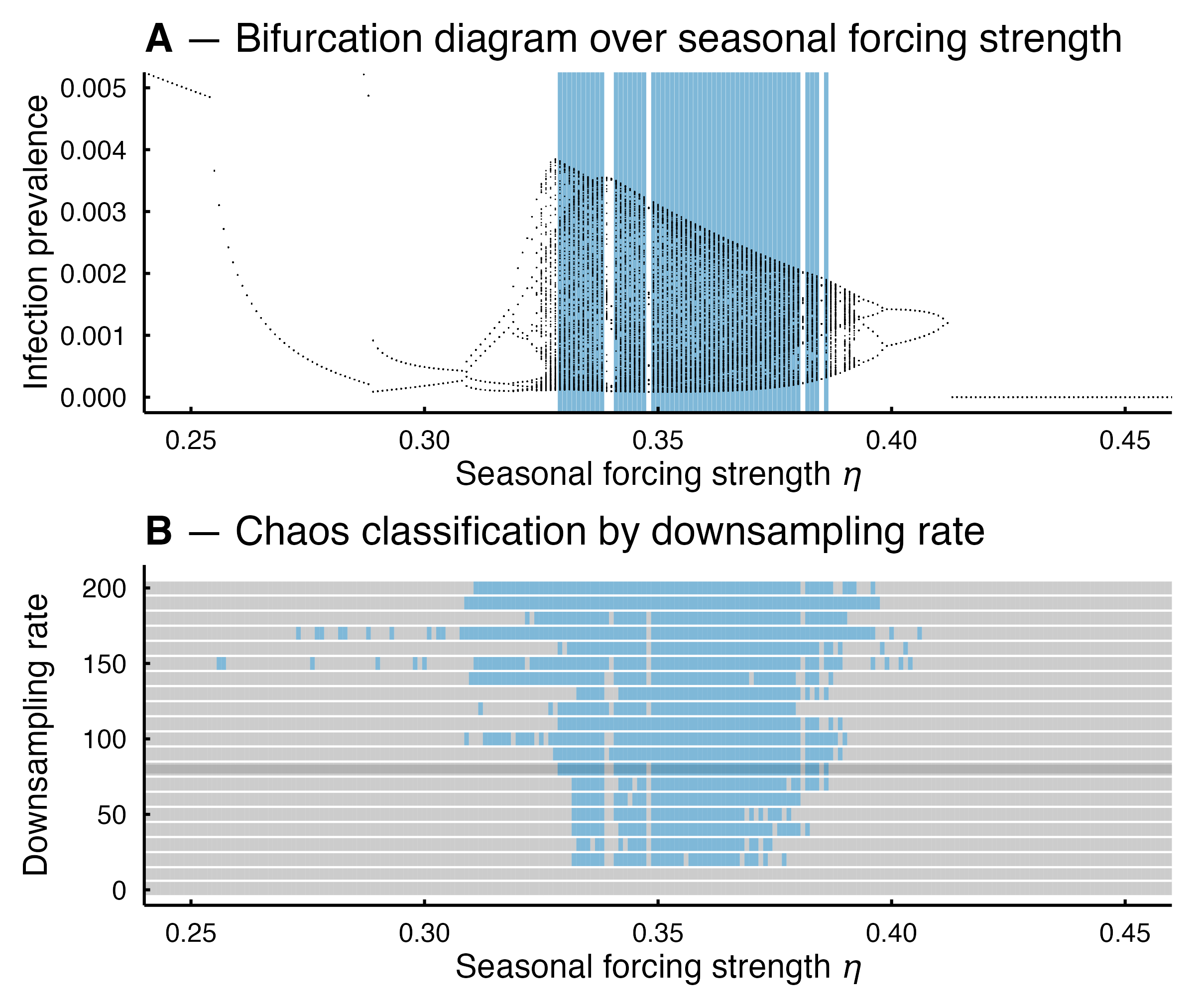}}
     \caption{ \textbf{A}: Bifurcation diagram over seasonality strength $\eta$, with effective decay rate fixed at $r = 0.018 \text{ days}^{-1}$. Period-doubling bifurcations are evident before chaotic behaviour emerges. Blue shading indicates where we have classified dynamics as chaotic using the Melbourne 0-1 test applied to the infection prevalence time-series downsampled at a rate of 80 days. \textbf{B}: Sensitivity analysis of the downsampling rate of the infection prevalence time-series to identify chaos. Blue shaded regions indicate where the Melbourne 0-1 test for chaos returned a value greater than $0.5$ for that combination of seasonality strength $\eta$ and downsampling rate. The row indicating the 80 day downsampling rate used in the manuscript is highlighted. }
    \label{figure-supp-chaos-downsample}
\end{figure}

\begin{figure}[H]
    \centering
    \makebox[\textwidth][c]{\includegraphics[width=13cm]{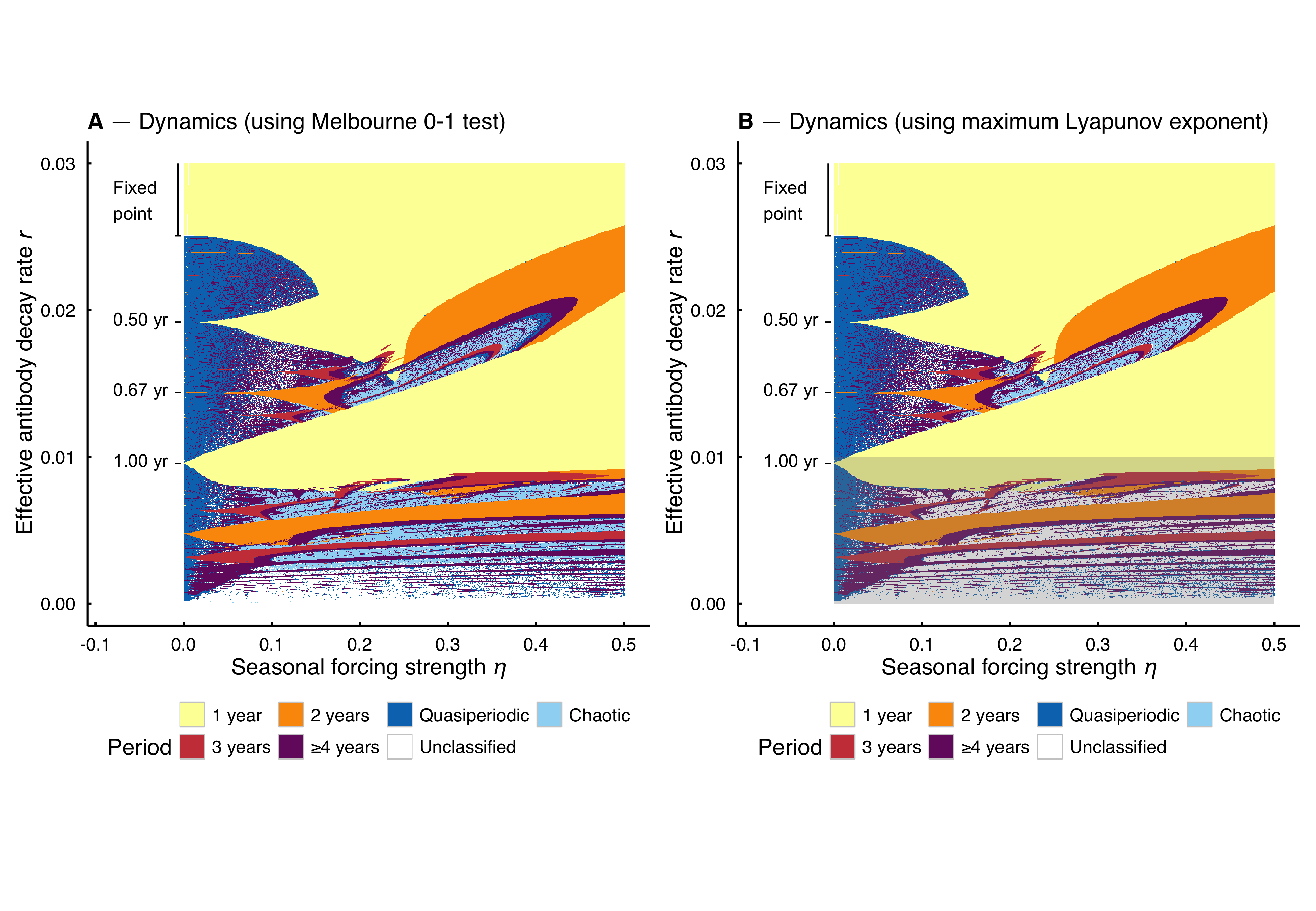}}
     \caption{ Comparison of chaos classification using the Melbourne 0-1 test compared to the numerically derived maximum Lyapunov exponent. \textbf{A}. Results as in Figure \ref{figure-grid-seasonality}B, using the Melbourne 0-1 test. \textbf{B}. Results using the numerically maximum Lyapunov exponent as a determinant of chaos. Dynamics which yielded Lyapunov exponents above a threshold $10^{-4}$ were classified as chaotic. Due to numerical instability, the maximum Lyapunov exponent could not reliably be calculated at low effective antibody decay rates. Hence, values were not calculated for $r < 0.01$, and this region is masked in light grey. }
    \label{figure-supp-chaos-lyapunov}
\end{figure}

Solutions where seasonal forcing is present can also be classified as chaotic or quasiperiodic. We define a (non-periodic) model solution to be chaotic where a value greater than $0.5$ is returned by the Melbourne 0-1 test \parencite{Datseris2018-kp}, with this test performed across infection prevalence sampled once every 80 days following burn-in. This value of 80 days was visually identified as achieving an acceptable classification performance (Appendix Figure \ref{figure-supp-chaos-downsample}). As a sensitivity analysis, we also consider the use of the numerically derived maximum Lyapunov exponent to identify chaos \parencite{Datseris2018-kp}, which generally agrees with the results of the Melbourne 0-1 test although suggests that a slightly larger region of parameter space may be chaotic (Appendix Figure \ref{figure-supp-chaos-lyapunov}). However, given that we observe numerical instability of the maximum Lyapunov exponent calculation for low effective antibody decay rates, the Melbourne 0-1 test is used as the primary test of chaos for this work.

Dynamics which are identified as not chaotic or periodic may be classified as quasiperiodic if we have at least two samples of $p_i$ but the value of $\bar{p}$ is not a multiple or sub-multiple of 365 days (as defined above). Model solutions which do not fulfil one of these listed criteria will remain unclassified, which may occur where we have particularly long transient periods that prevent a periodic or quasiperiodic solution from being identified or the chaos test returns a false-negative result.

\begin{figure}[H]
    \centering
    \makebox[\textwidth][c]{\includegraphics[width=10cm]{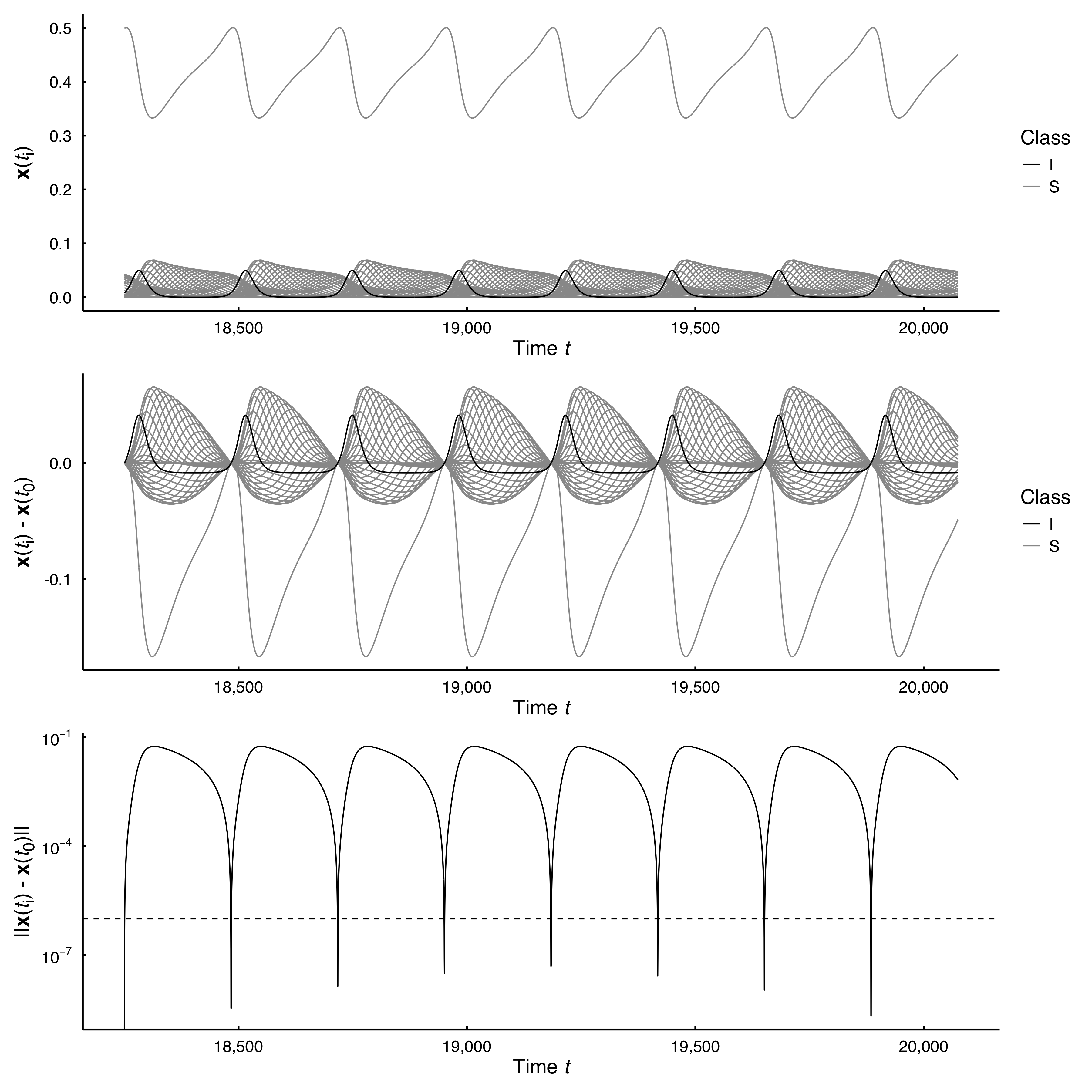}}
     \caption{Illustration of the condition used to determine period occurrence. Top: The distance $\mathbf{x}(t_i) - \mathbf{x}(t_0)$ across each compartment making up the state $\mathbf{x}$. Trajectories are illustrated following burn-in. Bottom: The Euclidean distance $||\mathbf{x}(t_i) - \mathbf{x}(t_0)||$, with a horizontal dashed line indicating the threshold for period occurrence ($\epsilon = 10^{-6}$), noting that the y-axis is on a log-scale.}
    \label{figure-ex-period-alg}
\end{figure}

\section*{Appendix D}

\textbf{Circular statistics for seasonal bias calculations}

We calculate circular statistics of mean and variance weighted by the daily infection incidence time-series over time $t$ (following burn-in) $\text{inc}(t)$ (where the calculation of incidence is described in the numerical methodology). We provide an illustration of this process in Appendix Figure \ref{figure-circular-mean} and detail the calculations in the following. To begin, we show how a time-point $t$ may be represented as a vector on the unit circle (where this circle represents the year, or 365 days, and corresponds to the phase and period of the seasonal forcing). We may calculate the time-of-year $t_\text{mod}$ of $t$ in days between 0 and 364:
\begin{equation*}
    t_\text{mod} = \text{mod}(t, 365).
\end{equation*}
This time-of-year corresponds to an angle $\theta$ counter-clockwise around the unit circle (Appendix Figure \ref{figure-circular-mean}):
\begin{equation*}
    \theta = \frac{2\pi t_\text{mod}}{365}.
\end{equation*}
And a corresponding vector $(x, y)$ along the unit circle:
\begin{equation*}
    x = \cos(\theta),\quad y = \sin(\theta),
\end{equation*}
or equivalently:
\begin{equation*}
    x = \cos\left(\frac{2\pi}{365}\text{mod}(t, 365)\right),\quad y = \sin\left(\frac{2\pi}{365}\text{mod}(t, 365)\right).
\end{equation*}

Using the above definition, we may calculate the mean vector $(\bar{x}, \bar{y})$ of time-of-infection, weighting by the daily infection incidence $\text{inc}(t)$:
\begin{align*}
    \bar{x} = \frac{\sum_{t\in T}\text{inc}(t) \cos(\frac{2\pi}{365} \text{mod}(t, 365))}{\sum_{t \in T} \text{inc}(t)}\\
    \bar{y} = \frac{\sum_{t\in T}\text{inc}(t) \sin(\frac{2\pi}{365} \text{mod}(t, 365))}{\sum_{t \in T} \text{inc}(t)}
\end{align*}
where $T$ is the sequence in days between the start of the post-burn-in period and the end of our numerical solution (where these must both be multiples of 365 such that only full years are included). We then calculate the circular mean day of year of infection $\bar{d}$ as:
\begin{align*}
    \bar{\theta} &= \begin{dcases}
        \text{atan2}(y = \bar{y}, x = \bar{x}) & \bar{y} \geq 0\\
        \text{atan2}(y = \bar{y}, x = \bar{x}) + 2\pi & \bar{y} < 0\\
    \end{dcases}\\[1em]
    \bar{d} &= \frac{365}{2 \pi}\bar{\theta},
\end{align*}
and the circular variance as one minus the magnitude of the mean vector:
\begin{equation*}
    v = 1 - \sqrt{\bar{x}^2 + \bar{y}^2}
\end{equation*}
which takes a value of zero where the timing of infection incidence is dispersed uniformly across the year and of one where all infections occur on the same day of the year.

\begin{figure}[H]
    \centering
    \makebox[\textwidth][c]{\includegraphics[width=12cm]{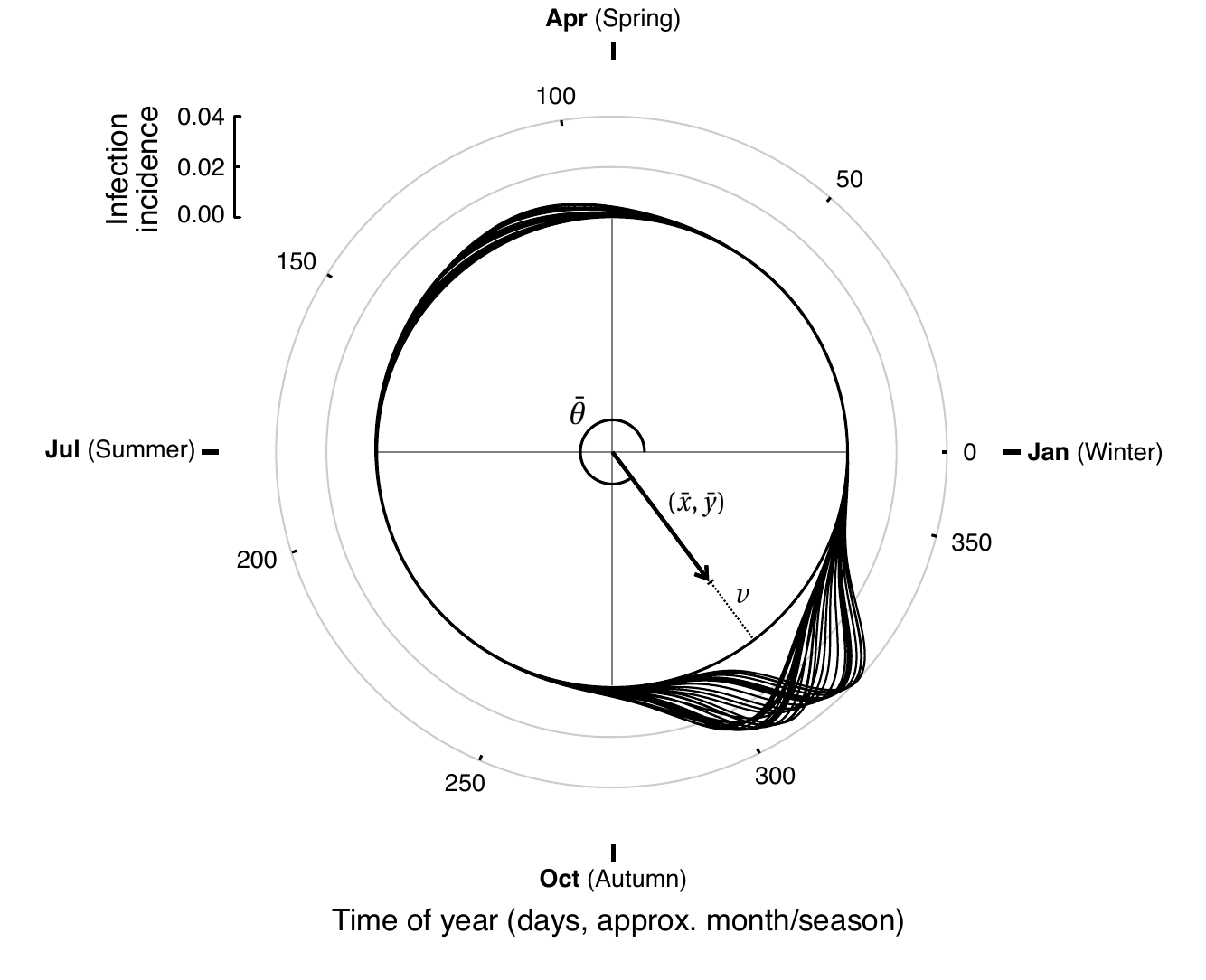}}
     \caption{Exemplar calculation of the circular mean and variance. Infection incidence time-series (following burn-in) is displayed around the circle by time of year $t_\text{mod}$ for $r = 0.06, \eta = 0.37$ (all other parameters as in Table 1). We have a mean vector of $(\bar{x}, \bar{y}) = (0.417 ,-0.549)$, corresponding to an angle of $\theta = 5.37$ and mean day of infection $\bar{d} = 312$. The circular variance is $v = 0.311$. }
    \label{figure-circular-mean}
\end{figure}

\section*{Appendix E}

\textbf{Parameter sensitivity analyses}

To assess the sensitivity of the bifurcation structure we identify in Figure \ref{figure-grid-seasonality}, we consider a number of model variations in Appendix Figure \ref{figure-supp-seasonality-sensitivity}. For Appendix Figures \ref{figure-supp-seasonality-sensitivity}B--G, we consider variations of parameters that affect the dynamics of antibody-mediated immunity. We find that changes that decrease dispersion in this process (i.e.\ increasing the Hill coefficient $h$ from $8$ to $12$ or narrowing the post-infection antibody distribution from $\text{N}(6, 0.5)$ to $\text{N}(6, 0.1)$) tend to preserve the overall bifurcation structure, while changes that increase dispersion (i.e.\ decreasing the Hill coefficient $h$ to $4$ or widening the post-infection antibody distribution to  $\text{N}(6, 1.0)$) result in a reduced complexity of the resultant bifurcation structure. result in large regions of period-1 dynamics. Decreasing the antibody-mediated protection mid-point $c_\text{mid}$ from 8 to 4 increases the duration of immune protection and in turn preserves the qualitative structure of the bifurcation diagram, while increasing $c_\text{mid}$ to 16 (reducing the duration of immune protection) also results in large regions of period-1 dynamics.

For Appendix Figures \ref{figure-supp-seasonality-sensitivity}H and \ref{figure-supp-seasonality-sensitivity}I, we consider increasing the degree of antibody level stratification $k$ from $32$ to $64$ and $128$ respectively. The degree of stratification $k$ mediates the dispersion in the antibody decay process (Appendix A), and higher values of $k$ may be appropriate where a lower degree of dispersion in the decay process is expected across the population. We find that the general structure of the bifurcation diagram is preserved, though with a vertical dilation similar to the other sensitivity analyses that produce a reduction in dispersion across the dynamics of antibody-mediated immunity (e.g. Appendix Figures \ref{figure-supp-seasonality-sensitivity}B and D).

\begin{figure}[H]
    \centering
    \makebox[\textwidth][c]{\includegraphics[width=17cm]{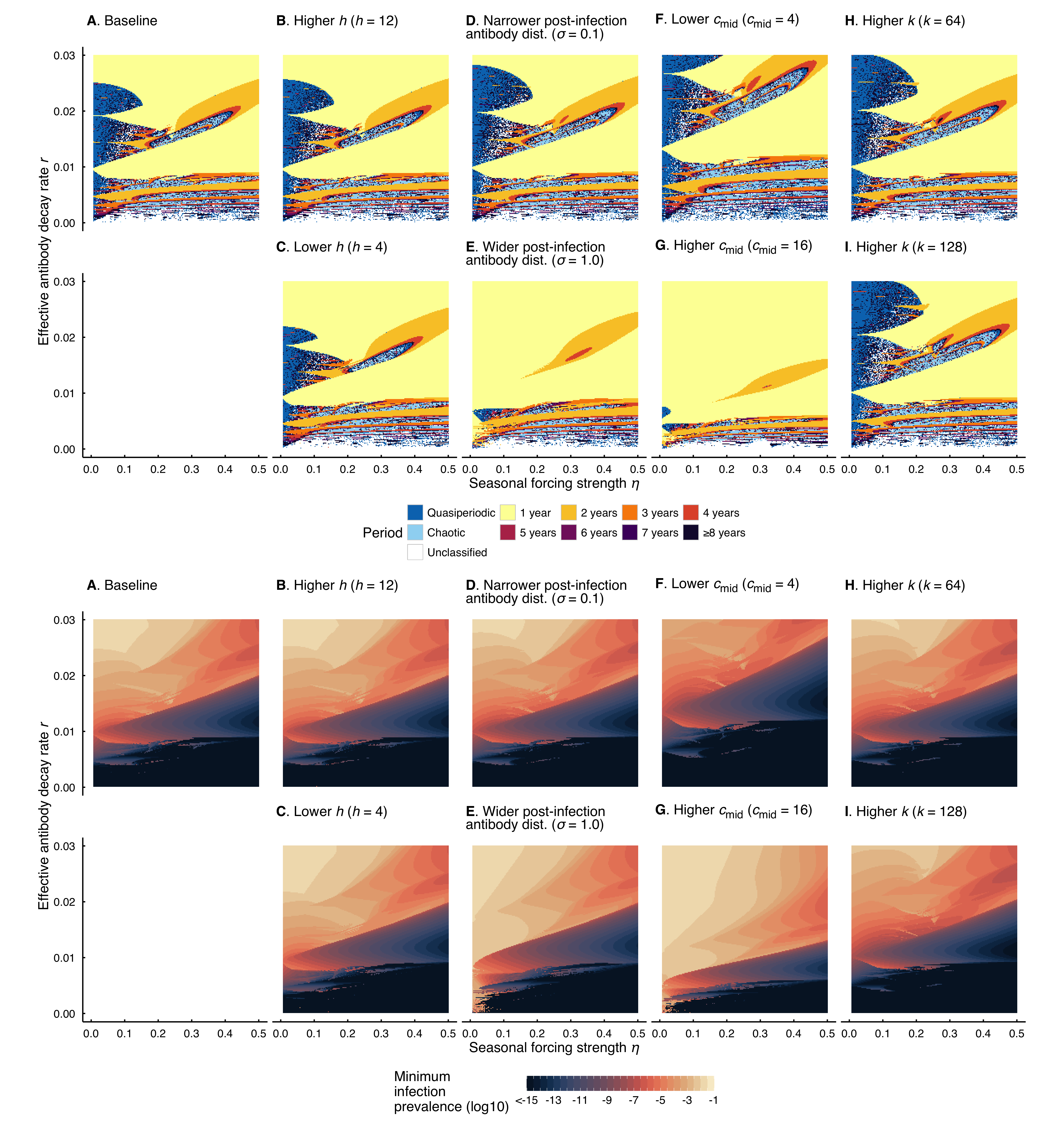}}
     \caption{ Sensitivity analysis of the dynamical results in Figure \ref{figure-grid-seasonality} across varied model parameters. Details on model parametrisation are provided in the text. }
    \label{figure-supp-seasonality-sensitivity}
\end{figure}

\section*{Appendix F}

\textbf{Stratified infectious class sensitivity analyses}

We also consider sensitivity analyses of Figure \ref{figure-grid-seasonality} where we use the system with additional stratification across the infectious class (Appendix B). This allows us to consider the behaviour of the system where antibody boosting occurs (as described in Appendix B) and where variation in the recovery rate occurs according to pre-existing antibody levels.

\begin{figure}[H]
    \centering
    \makebox[\textwidth][c]{\includegraphics[width=14cm]{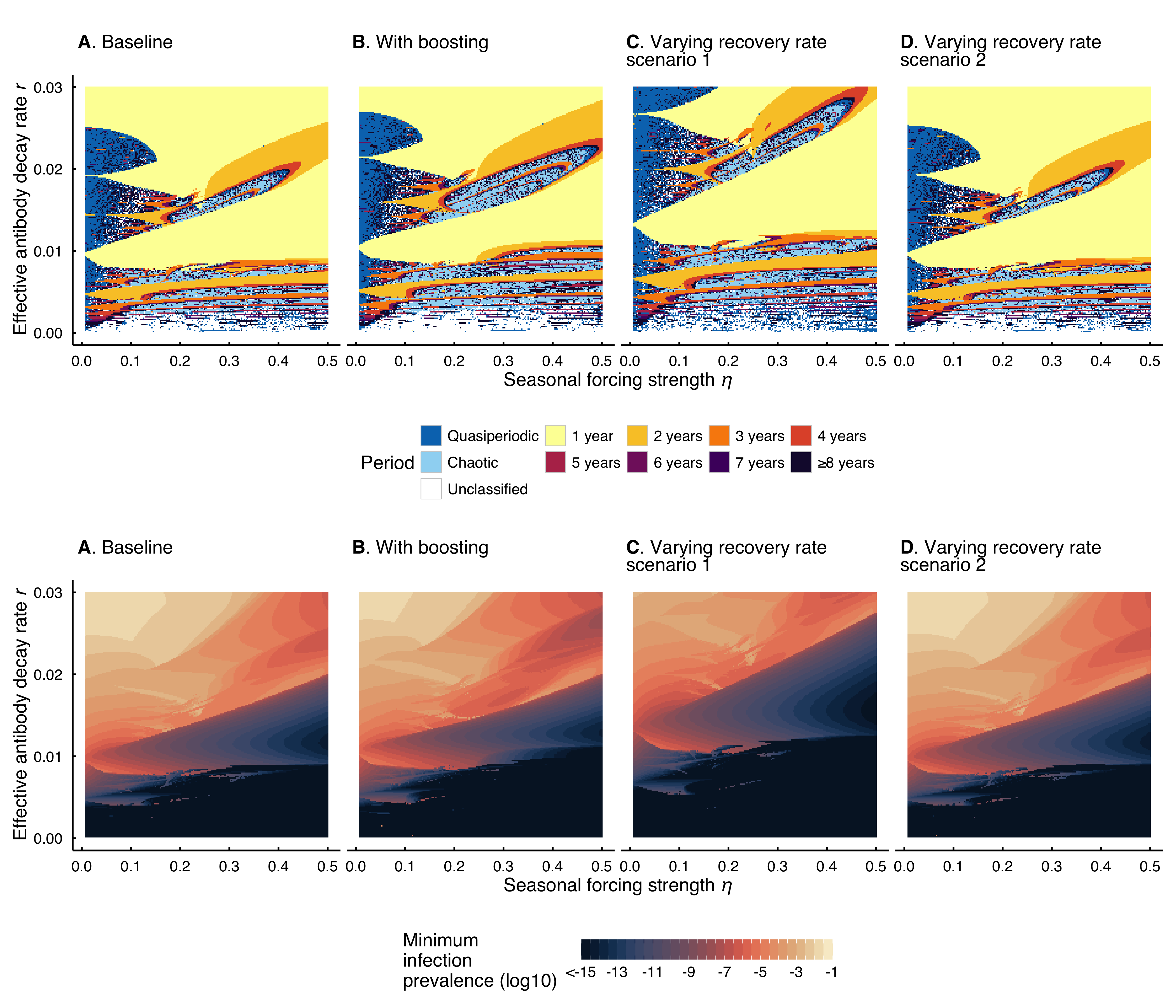}}
     \caption{ Sensitivity analysis of the dynamical results in Figure \ref{figure-grid-seasonality} where we also stratify the infectious class. \textbf{A}. Baseline results. \textbf{B}. Results from the model with boosting as described in Appendix B. \textbf{C}--\textbf{D}. Results from a model with varying recovery rate according to pre-existing antibody level. The two scenarios are described in the text.}
    \label{figure-supp-seasonality-sensitivity-strat}
\end{figure}

For the scenario with antibody boosting (Appendix Figure \ref{figure-supp-seasonality-sensitivity-strat}B), we find that the inclusion of antibody boosting has relatively little effect on the overall qualitative structure of the bifurcation diagram, although it does expand the region of the parameter space that yields chaos.

In Appendix Figures \ref{figure-supp-seasonality-sensitivity-strat}C--D, we consider a version of the model where the infectious class is also stratified by antibody level, as presented in Appendix B. However, rather than including antibody boosting, we explore the effect of varying the recovery rate according to pre-infection antibody levels. For this, we let the recovery rate for an individual in stratum $i$ be:
\begin{equation*}
    \gamma_i = \gamma + \frac{c_i^h}{c_\text{mid-rec}^h+c_i^h},
\end{equation*}
where $\gamma = 0.25$ is the baseline rate of recovery, $h = 8$ is the Hill coefficient, and $c_\text{mid-rec}$ is the mid-point of the Hill curve. For low effective antibody levels, recovery occurs at a rate of 0.25 (mean 4 days). For high effective antibody levels, recovery occurs at a rate of 1.25 (mean 0.8 days).

We specify two scenarios for $c_\text{mid-rec}$. In scenario 1, we take $c_\text{mid-rec} = 4$ such that the recovery rate curve has a mid-point at a lower effective antibody level than the protection curve (see Appendix Figure \ref{figure-supp-gamma-scenarios}). In scenario 2, we take $c_\text{mid-rec} = 16$ such that recovery rate curve has a mid-point at a higher effective antibody level than the protection curve. These two scenarios demonstrate that the epidemiological-scale dynamics are dominated by the protection effect that acts at the lower effective antibody level. That is, in scenario 1, antibody-mediated recovery rate supersedes the effects of antibody-mediated protection against infection (Appendix Figure \ref{figure-supp-seasonality-sensitivity-strat}C). This result is highly similar to reducing the midpoint of the protection against infection curve (as in Appendix Figure \ref{figure-supp-seasonality-sensitivity}F). In contrast, in scenario 2, antibody-mediated protection supersedes any effects of the antibody-mediated recovery rate, and results are highly similar to baseline (Appendix Figure \ref{figure-supp-seasonality-sensitivity-strat}D). As such, a model that captures only the antibody-mediated protection against infection (as in the main text) would be sufficient to approximate the potential effects of an antibody-mediated recovery rate.

\begin{figure}[H]
    \centering
    \makebox[\textwidth][c]{\includegraphics[width=10cm]{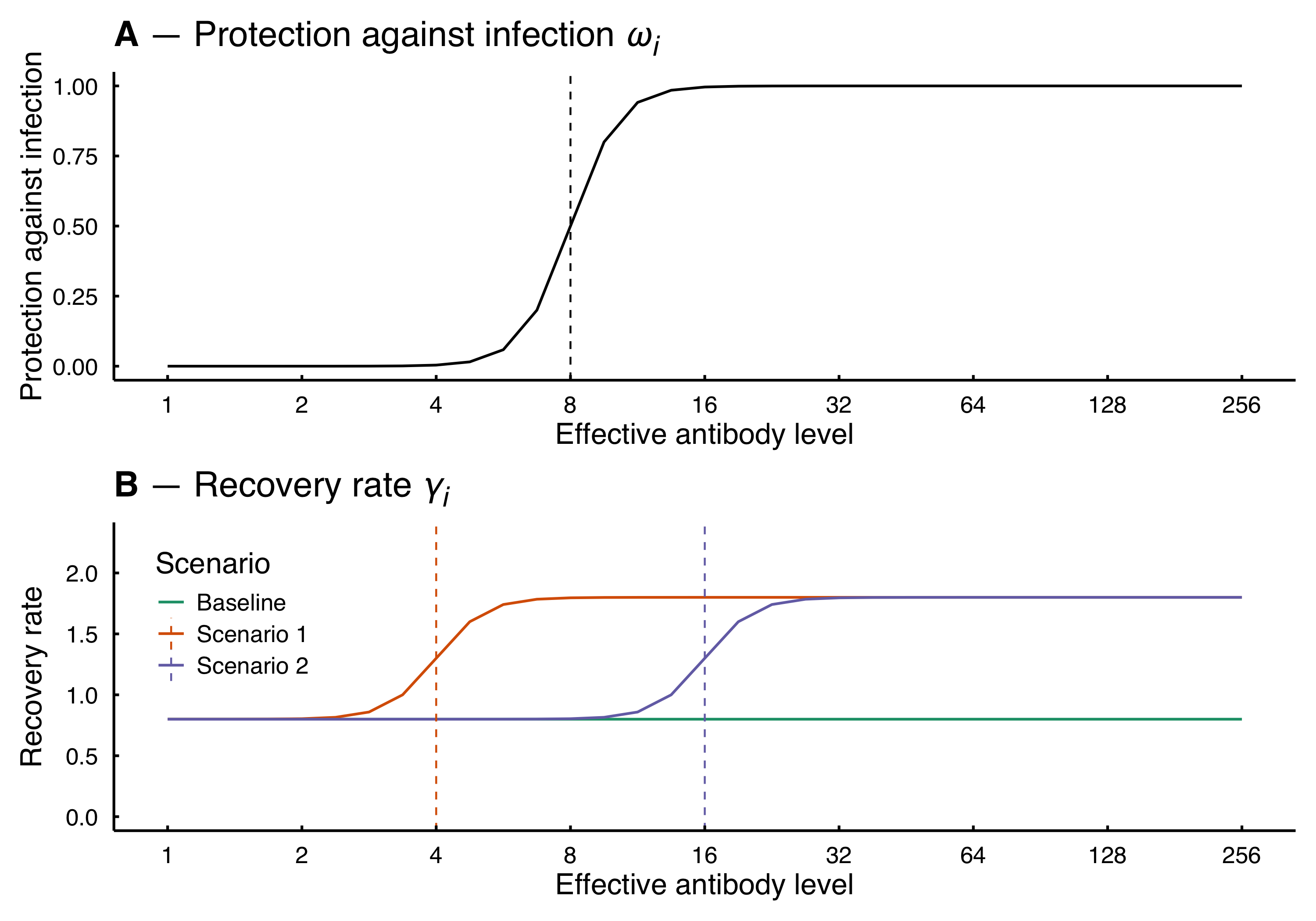}}
     \caption{ \textbf{A}. The relationship between effective antibody level and protection against infection $\omega_i$. \textbf{B}. The relationship between effective antibody level and recovery rate $\gamma_i$ at baseline and the two scenarios considered in Appendix Figure \ref{figure-supp-seasonality-sensitivity-strat}. }
    \label{figure-supp-gamma-scenarios}
\end{figure}

\section*{Appendix G}

\textbf{Examination of the seasonally-forced \textit{SIRS} model}

To compare our approach to the classical epidemiological framework of the \textit{SIRS} model \parencite{Berger1976-vf}, we apply the numerical analysis methodology used in Figure \ref{figure-grid-seasonality} to this simple three-compartment model. We consider the model without demography and with a constant decay rate:
\begin{align*}
    \dv{S}{t} &= -\beta(t)SI+\sigma R \\
    \dv{I}{t} &= \beta(t)SI - \gamma I \\
    \dv{R}{t} &= \gamma I - \sigma R
\end{align*}
where $\beta(t)$ and $\gamma$ are specified as in the main text. For this model, we use the immunity waning rate $\sigma$ as a stand-in for the effective antibody decay rate $r$. In Appendix Figure \ref{figure-supp-seasonality-SIRS}, we vary the seasonal forcing strength $\eta$ and the rate of waning $\sigma$, numerically solve the system across $1,250$ years and classify the dynamics observed across the final $250$ years. In Appendix Figure \ref{figure-supp-seasonality-SIRS}A, we see that the bifurcation structure is dissimilar to the results for our antibody-stratified model. In particular, Arnold tongues are less visually apparent. Notably, a similar period-doubling route to chaos appears to be present. The minimum infection prevalence across the parameter space is similar to what we observe in Figure \ref{figure-grid-seasonality}, although prevalence is much higher for combinations of low forcing strength and low waning rate (since the model cannot produce periodic solutions in the absence of forcing). The presence of quasiperiodicity was detected by the numerical classification algorithm but visual checks showed these were false positives (and as such no quasiperiodicity is depicted in Appendix Figure \ref{figure-supp-seasonality-SIRS}A). These false positives were likely due to the lower dimensionality of this system, and as such a full numerical study of the seasonally forced \textit{SIRS} system would need to consider a methodology more appropriate for the size of the system. These results are similar to what we observe in Appendix Figure \ref{figure-supp-seasonality-sensitivity} where parameters in the full model are varied such that dispersion in the dynamics of antibody-mediated immunity is high. As such, where we expect high dispersion in the antibody-mediated protection, the \textit{SIRS} model may be a suitable stand-in for our model. However, where low dispersion is expected, the \textit{SIRS} model would fail to capture the full range of potential dynamical behaviours that we observe in Figure \ref{figure-grid-seasonality}.

\begin{figure}[H]
    \centering
    \makebox[\textwidth][c]{\includegraphics[width=14cm]{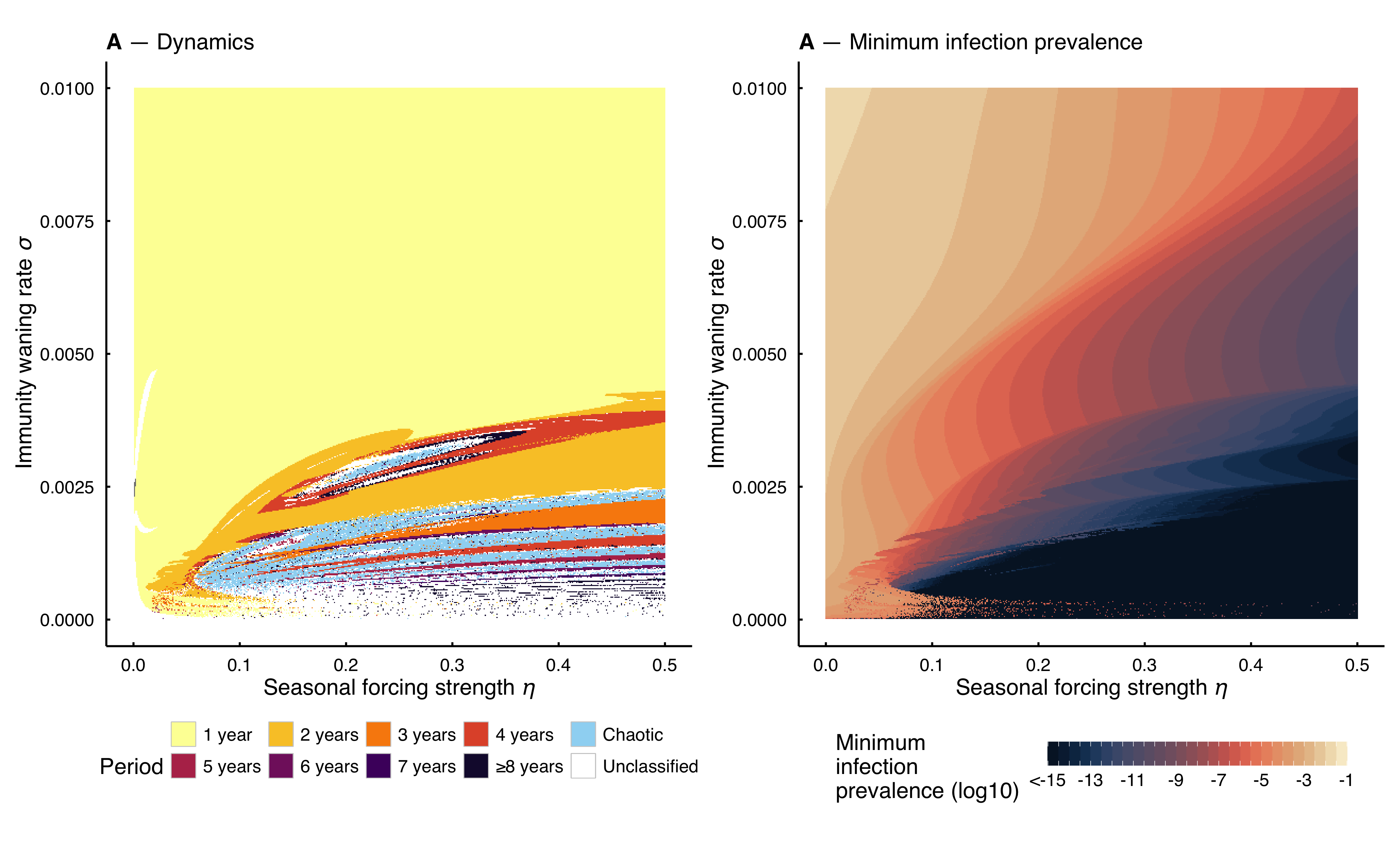}}
     \caption{ Bifurcation analysis of the SIRS model using the same numerical methodology applied in Figure \ref{figure-grid-seasonality}. See the text for full details. }
    \label{figure-supp-seasonality-SIRS}
\end{figure}

\printbibliography
\end{refsection}

\end{document}